\documentclass{aa}
 
\begin{document}
 
\thesaurus{03 (11.09.5; 11.09.1 IC1613; 11.11.1; 11.12.1; 09.08.1; 09.02.1)}

\title{Kinematics of the ionized gas in the Local Group irregular galaxy
\object{IC 1613}}

\author{M. Valdez-Guti\'errez \inst{1}, 
        M. Rosado \inst{2},
        L. Georgiev \inst{2},
        J. Borissova \inst{3}
        \and
        R. Kurtev \inst{4}}

\institute{Instituto Nacional de Astrof\'\i sica, Optica y Electr\'onica,
           Calle Luis Enrique Erro 1, 72840 Tonantzintla, Puebla, M\'exico
           \and 
           Instituto de Astronom\'\i a, Universidad Nacional Aut\'onoma de
           M\'exico, M\'exico
           \and
           Institute of Astronomy, Bulgarian Academy of Sciences
           and Isaac Newton Institute of Chile Bulgarian Branch,
           72 Tsarigradsko chauss\`ee, BG --1784 Sofia, Bulgaria
           \and
           Department of Astronomy, Sofia University 
           and Isaac Newton Institute of Chile Bulgarian Branch, 
           BG --1164 Sofia, Bulgaria 
           }

\offprints{M. Valdez-Guti\'errez}
\mail{mago@inaoep.mx}
 
\date{Received 7 June 2000   / Accepted 20 September 2000}

\titlerunning{Kinematics in \object{IC 1613}}
\authorrunning{M. Valdez-Guti\'errez et al.}

\maketitle
 
\begin{abstract}
We present H$\alpha$ and [\ion{S}{ii}] observations for the Local Group
irregular galaxy \object{IC 1613} using the PUMA scanning Fabry--Perot
interferometer.
Our goal is to analyze the kinematics of the ionized gas in the complex sample
of superbubbles located in the whole extension of our field (10\arcmin ), which
includes most of the optical emission of this galaxy, and to study the
inter-relationship between young stellar associations and nebulae based on a
previous study that we have made on the stellar associations of the central
region of this galaxy. 
The ionized gas in this galaxy is distributed in classical \ion{H}{ii} regions
and in a series of superbubbles (also called giant shells) covering a large
fraction of the optical extent of the galaxy.

We present a catalog of kinematical properties of both the \ion{H}{ii} regions
of this galaxy and the superbubbles.
We have also compared the kinematics of the ionized gas in \ion{H}{ii}
regions to search for possible dynamic differences between neutral and
ionized gas. 
   
\keywords{Galaxies: irregular -- Galaxies: individual: \hbox{\object{IC 1613} }
-- Galaxies: kinematics and dynamics -- Local Group --  (ISM): HII regions 
-- ISM: bubbles}

\end{abstract}

\section{Introduction}
 
Irregular galaxies constitute a substantial fraction of all galaxies. They are
also dominant among the actively star-forming galaxies. Because they lack 
spiral density waves, they are excellent laboratories to examine
the star formation process in alternative scenarios. Also, they can provide
important clues about the mechanisms of formation of
``bubbles'' (the ring-shaped nebulae with diameters, D $\la$ 50 pc),
``superbubbles'' or ``giant shells''(ring-shaped structures with D $\la$ 400 
pc) and ``supershells'' (ring-shaped structures with D $\approx$ 1 kpc).
Massive stars can dramatically change the interstellar medium (ISM) of these
galaxies by means of supernova explosions and stellar winds. On the other hand,
the ISM content and energetics can have important effects on the subsequent
mechanisms of star formation. Indeed, the compressions produced by shocks formed
by supernova explosions or powerful stellar winds on the molecular clouds 
could trigger the formation of a second generation of stars at the edges of the
shells formed by these events (self-induced star formation).
 
The Local Group irregular galaxies are exceptional targets for the study of 
the inter-relationship between gas and stars because they are easily resolved.
This means that we could know the stellar content of nebular \ion{H}{ii} 
complexes and superbubbles and try to identify for relationships between the
interior stellar associations and the  dimensions and kinematics of the
superbubbles.
This kind of investigation has been performed for the most favorable case, 
the Large Magellanic Cloud (Rosado \cite{rosado86}; Oey \cite{oey96a}, 
\cite{oey96b}, amongst other references). 
Indeed, it is in this galaxy that a clear example of self--induced star 
formation has been found: the nebular complex N11
(Parker et al. \cite{parker96}; Rosado et al. \cite{rosado96}).

In this work, we study the nebulae and massive stellar content of the nearby
irregular galaxy: \object{IC 1613}. This galaxy belongs to the Local Group 
and it is located at a distance of 725 kpc (Freedman \cite{freedman88a}, 
\cite{freedman88b}). 
Table \ref{properties} reviews some of its main  properties.
Because of its proximity, it has been the target of several studies.
Its stellar content has been analyzed by Baade (\cite{baade63}), Sandage
(\cite{sandage71}), Sandage \& Katem (\cite{sandage76}), Hodge (\cite{hodge78},
\cite{hodge80}), Freedman (\cite{freedman88a}, \cite{freedman88b}) and Georgiev
et al. (\cite{georgiev99}; hereafter Paper I). 
Several candidate Wolf-Rayet (WR) stars were detected but only one of them
was spectroscopically confirmed over the whole extent of this galaxy
(Azzopardi et al. \cite{azzopardi88}; d'Odorico \& Rosa \cite{d'odorico82};
Kingsburgh \& Barlow \cite{kingsburgh95}). 
This WR is a WO--type star embedded in a large nebular complex located at the 
southern region of this galaxy (Rosado et al. \cite{rosado00}). 
The \ion{H}{ii} regions of \object{IC 1613} have been studied by Sandage
(\cite{sandage71}), Hodge et al. (\cite{hodge90}; hereafter HLG), Price et
al. (\cite{price90}) and Hunter et al.  (\cite{hunter93}). 
The peculiarity of the kinematics of some of the superbubbles of ionized gas 
has been analyzed by Meaburn et al. (\cite{meaburn88}). 
Their work focused only on the NE region of the galaxy where bright \ion{H}{ii} 
complexes are easily detected. 
Other spectroscopic studies on the nebulae of \object{IC 1613} mainly examine the 
only SNR (d'Odorico et al. \cite{d'odorico80}; Peimbert et al. \cite{peimbert88}; 
Dickel et al. \cite{dickel85}; Goss \& Lozinskaya \cite{goss95}; Rosado et al.
\cite{rosado00}).
The distribution of the \ion{H}{i} gas in \object{IC 1613} has been analyzed by
Lake \& Skillman (\cite{lake89}). These authors have found that the \ion{H}{i}
content makes up to 20 percent of the total mass (see also Huchtmeier et al.
\cite{huchtmeier81}). 

In this work, the results of our kinematic analysis, making use of a scanning
Fabry-Perot interferometer (PUMA, Rosado et al. \cite{rosado95}), are presented.
We analyze the relationship between nebulae and the stellar associations 
studied in Paper I. 
The observations and data reductions are described in Section 2.
In the remaining sections we study the kinematic behavior of the \ion{H}{ii}
regions and superbubbles. In Section 3 we show the nebulosities of this 
galaxy in several emission lines. 
We also indicate how our photometric calibration has been done. 
In Section 4 we study the kinematics of HII regions and superbubbles in
\object{IC 1613}. 
We obtain the peak velocities and velocity dispersions of the \ion{H}{ii}
regions, as well as their stellar content. We also obtain the \ion{H}{ii}
luminosity function and diameter distribution. 
In  Section 5 we study the superbubbles found, obtaining their dimensions,
expansion velocities, H$\alpha$ fluxes and studying their stellar content. 
We also obtain some important physical parameters of the superbubbles and 
the superbubble diameter distributions. 
We analyse the possibility that the superbubbles were formed by the combined
action of stellar winds and supernovae explosions. 
In Section 6 we present the radial velocity field in our field of view,
obtained from diffuse ionized gas, \ion{H}{ii} regions and superbubbles.
We compare our trends with the results of  \ion{H}{i} observations. 
In Section 7 we present our conclusions.

\section{Observations and data reduction}

The observations were carried out during the nights of December 5 and 6, 1996 
and November 15, 1998 with the UNAM Scanning Fabry-Perot 
interferometer  PUMA attached to the f/7.9 Ritchey-Chretien focus 
of the 2.1 m telescope at the Observatorio Astron\'omico Nacional at San 
Pedro M\'artir, B.C., M\'exico.

The PUMA setup was composed of a scanning Fabry-Perot interferometer (SFPI),
a focal reducer with a f/3.95 camera, a filter wheel, a calibration system
and a Tektronix 1K$\times$1K CCD detector (Rosado et al. \cite{rosado95}).
Table \ref{observations} reports the main characteristics of the instrumental
setup.

We obtained a series of direct images in the lines of H$\alpha$, [\ion{S}{ii}]
($\lambda$ 6717 and 6731 \AA) and [\ion{O}{iii}] ($\lambda$ 5007 \AA) using the
PUMA in its direct image mode (without the SFPI). 
The exposure times of each of the direct images were 60 s, 60 s and 300 s for 
the H$\alpha$, [\ion{S}{ii}] and [\ion{O}{iii}], respectively . The CCD reading
was binned 2$\times$2, giving a pixel size of 1\farcs 17 (equivalent to 4.21
pc at the adopted distance) and a field of view covering 
10$\arcmin$ including the NE region and several other interesting regions of
\object{IC 1613}.
 
In addition, we obtained scanning FP interferometric data cubes in the
lines of H$\alpha$ and [\ion{S}{ii}] ($\lambda$ 6717 and 6731 \AA).
The H$\alpha$ cubes share the same scale and field of view as the direct images.
Given that the [\ion{S}{ii}] intensities were lower than the H$\alpha$
intensities, we binned 4$\times$4 the CCD reading of the [\ion{S}{ii}]
cubes and correspondingly, we obtained a pixel size of 2\farcs34.

The phase calibration was made by taking data cubes of a calibration lamp
before and after the galaxy exposure. We used a hydrogen lamp for
calibrating the H$\alpha$ nebular cubes and a neon lamp for calibrating the
[\ion{S}{ii}] nebular cubes.
Since the redshift of this galaxy is not high, phase corrections, due to the
difference in wavelength between the galaxy and the calibration line, are not
required.

The FP data cubes are composed of 48 steps, with integration times of 60 seconds
per step. In this form, the final data cube dimensions in H$\alpha$ and
[\ion{S}{ii}] were 512$\times$512$\times$48 and 256$\times$256$\times$48,
respectively.
For the H$\alpha$ observations we obtained 4 data cubes that were summed
giving a total exposure time of 3.2 hours.
For the [\ion{S}{ii}] observations we obtained 3 data cubes that also were 
summed, with a total exposure time of 2.4 hours.

Reductions were carried out using the software package CIGALE (Le Coarer et
al. \cite{lecoarer93}).
Our data cubes in H$\alpha$ are contaminated with line-sky emission such as
the geocoronal H$\alpha$ line and the lines of OH at 6577.18 \AA\ and 6577.38
\AA, respectively. 
The night-sky lines fell well outside the main emission of \object{IC 1613}, due
to the systemic velocity of this galaxy.
These night-sky lines were subtracted using an interactive routine of CIGALE. 
The data cubes in the [\ion{S}{ii}] lines are less contaminated by line-sky
emission; however, the S/N ratio is not as good as that of the H$\alpha$ cubes 
and so will be used only for comparison.

To extract the kinematic information from the FP data cubes we proceeded in 
different ways for the \ion{H}{ii} regions and for the superbubbles. In the case 
of the \ion{H}{ii} regions, we obtained only one radial velocity profile at 
H$\alpha$, integrated over a box containing the whole \ion{H}{ii} region. In the 
case of the superbubbles, we proceed in a different way in order to seek for 
internal motions revealed by line splitting of the velocity profiles. We 
obtained the radial velocity field of the superbubble  by dividing the 
superbubble dimensions into several small boxes over which we obtained several 
radial velocity profiles, both in H$\alpha$ and [\ion{S}{ii}].

The radial velocity profiles were fitted by Gaussian functions after 
deconvolution by the instrumental function (an Airy function).
We constructed a map of the monochromatic (continuum) emission by integration of
the radial velocity profile of each pixel, up (down) to a certain fraction of
the peak.
This map enabled us to separate the monochromatic emission from the continuum
(Le Coarer et al. \cite{lecoarer93}).
To get the radial velocity map of our observed field, we found the central 
velocity in the radial velocity profile of each pixel, that has a level above 
2 $\sigma$ the value of the standard background emission, to ensure galaxy 
membership.

\section{Morphology of the nebulae in \object{IC 1613}}

Figure \ref{direct} shows the emission of \object{IC 1613} in the lines of
H$\alpha$, [\ion{S}{ii}] and [\ion{O}{iii}] as obtained from our direct images.
The photometrical center of \object{IC 1613} is located at RA(1950)=01:02:19.6
and Dec(1950)=+01:51:56.1 (Mateo \cite{mateo98}).
The most active region of \object{IC 1613} is located 3$\arcmin$ (or 663 pc)
North-East from the photometric center. Hereafter, we will call this quadrant
the NE region.
Bright \ion{H}{ii} regions, bubbles and superbubbles are situated in the NE
region.
Numerous \ion{H}{ii} regions and larger structures, some of them ring--shaped,
are superimposed over the extent of the chaotic body of the galaxy.
Except for the bright network of superbubbles located in the NE region of the
galaxy, the nebulae are weaker in the [\ion{S}{ii}] lines and they do not show
any appreciable [\ion{O}{iii}] emission.  

In the South of this rich sector of the galaxy (at 3\farcm8 or 838 pc from the
center), there is a giant \ion{H}{ii} region (Sandage 3). Southwest of
the center of the galaxy at 2\farcm4 (535 pc) we located the \ion{H}{ii}
region Sandage 2. 

\begin{figure}[ht]
%\special{psfile=ms9979f1.ps angle=0 hscale=97 vscale=97 voffset=-717 
%hoffset=-130}
\vspace{19.5cm}
\caption{Top to bottom: Direct images of \object{IC 1613} in H$\alpha$, 
[\ion{S}{ii}] and [\ion{O}{iii}], respectively. The exposure time is 60 sec
for H$\alpha$ and [\ion{S}{ii}], while for [\ion{O}{iii}] it is 300 sec (see
text for details). The scale and the orientation are indicated in the upper
panel.}
\label{direct}
\end{figure}

\subsection{The \ion{H}{ii} region content}

The \ion{H}{ii} regions of \object{IC 1613} have been catalogued by Sandage
(\cite {sandage71}), Lequeux et al. (\cite{lequeux87}), Price et al.
(\cite{price90}) and HLG (who catalogued 77 \ion{H}{ii} regions, reporting 
their diameters and H$\alpha$ fluxes). 

Our observed field contains 44 diffuse nebulae according to HLG. The limits of
each \ion{H}{ii} region or group of \ion{H}{ii} regions were defined by the
contour where the intensity of the H$\alpha$ emission falls to the
average local intensity of the diffuse background. However, in some cases, the
boundaries between diffuse \ion{H}{ii} regions are not obvious. In addition, we
were not able to resolve some regions. In these cases, we assign to them a
common boundary, and, consequently, only 30 regions were clearly defined.

\subsection{Photometric calibration}

\begin{figure}[ht]
\includegraphics{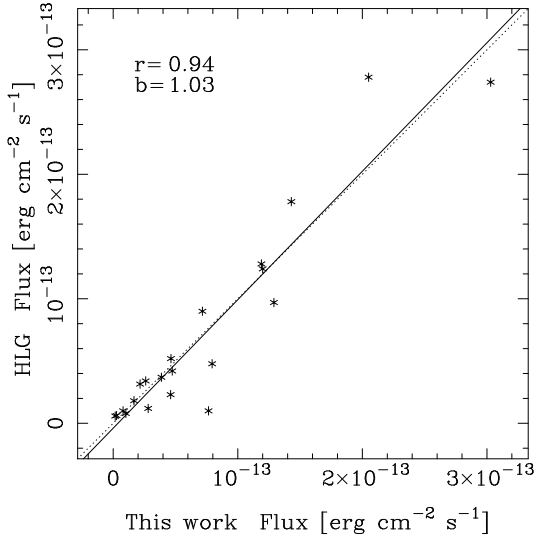}
\vspace{6.8cm}
\caption{Comparison between our flux determinations and HLG fluxes. The
calibrator HII regions were not used in the fitting. The values of the
correlation coefficient (r) and the slope (b) are given to the left.
The dotted line is our best fit, and the solid line is the diagonal.}
\label{comparison}
\end{figure}

We calibrate our monochromatic H$\alpha$ image in flux, using the measurements
of HLG. Eight isolated and round regions were selected from HLG lists 
(Table \ref{standard}).
For each region we measured the flux of our monochromatic H$\alpha$ image, using
apertures with the same equivalent area as in HLG. A least squares fit to the
data gives a conversion factor of 
1 count  s$^{-1}$ (in our H$\alpha$ image) = 6.26 10$^{-18}$ \mbox{erg cm$^{-2}$
s$^{-1}$}.
We preferred a conversion factor per unit area for easy calculations of the flux
in irregular HII regions and superbubbles.
Polygonal apertures were chosen such that they must contain all emission from
a given \ion{H}{ii} region or superbubble. Pixels within the apertures were
summed to generate the total \ion{H}{ii} region or superbubble plus background
counts.
The background emission was computed using apertures with the same area as 
the object, as close to the object as possible.
The total H$\alpha$ flux for every \ion{H}{ii} region and every superbubble was
obtained when the background was subtracted and the resulting counts multiplied
by the conversion factor.
H$\alpha$ fluxes are presented in column 9 of Table \ref{regiones} for the 
\ion{H}{ii} regions and column 6 of Table \ref{cat_shells} for the superbubbles. 
Figure \ref{comparison} shows the comparison between our measurements and HLG 
for the common HII regions not included as calibrators.

\begin{figure*}[htb]
\includegraphics{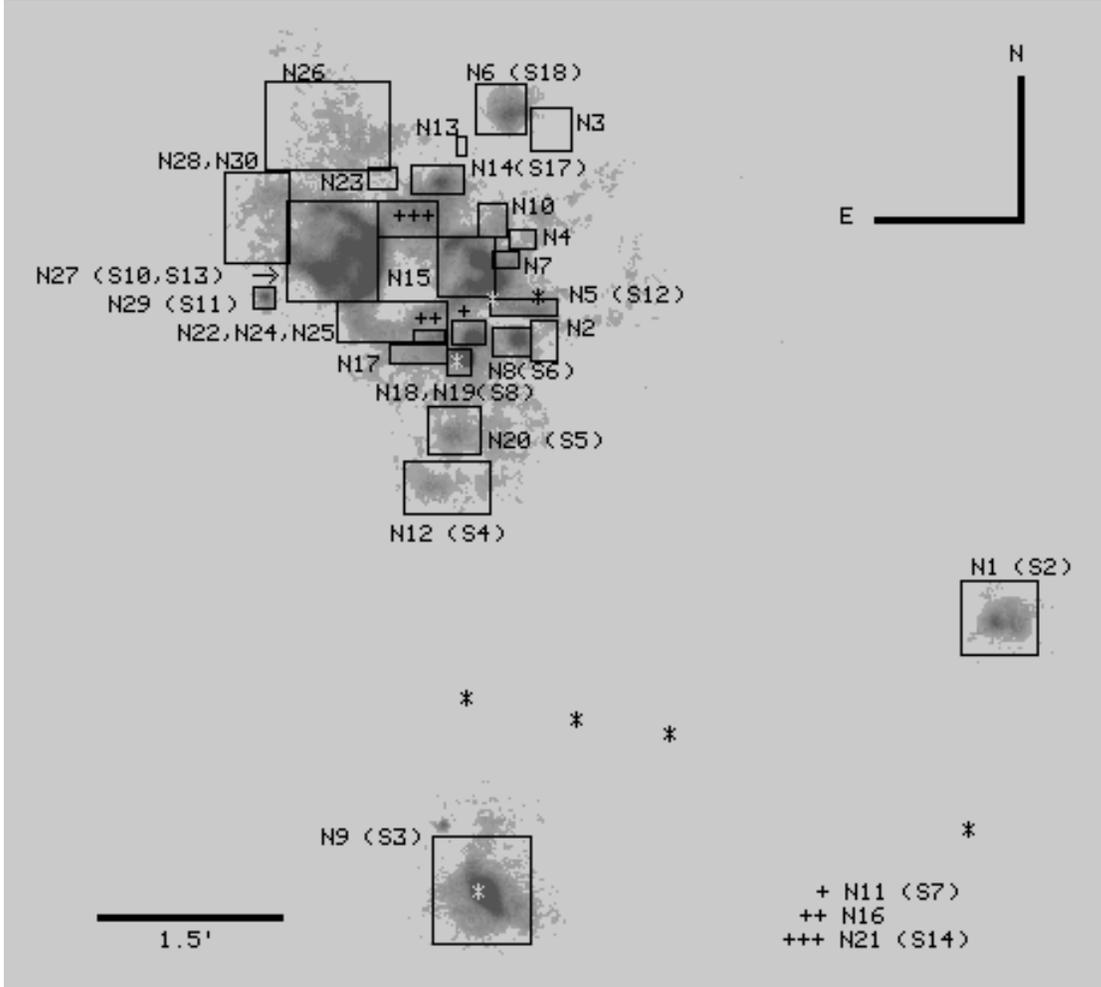}
\vspace{14.1cm}
\caption{H$\alpha$ monochromatic image of \object{IC 1613}. The background 
has been subtracted at $2\sigma$ its mean value in order to reveal 
the weaker features. The boxes correspond to places where the kinematical
analysis was performed. Sandage identifications have been quoted between
parentheses if relevant. The asterisks indicate positions for the eight WR star
candidates (Armandroff \& Massey \cite{armandroff85}). Some labels are
denoted with crosses, at the left bottom of the figure. The scale and 
orientation are indicated to the left and right, respectively. 
In this Figure the field covers approximately 9$\arcmin\times9\arcmin$.}
\label{box-vel}
\end{figure*}

\section{Kinematics of the \ion{H}{ii} regions in \object{IC 1613}}

Figure \ref{box-vel} shows the H$\alpha$ monochromatic image of 
\object{IC 1613} (continuum subtracted) obtained from our FP cubes using the 
method described in Section 2.
Superimposed on this image, we marked the rectangular boxes over which the 
radial velocity measurements of the observed \ion{H}{ii} regions were 
performed.
We obtained radial velocity profiles of these sources by integrating in those
rectangular boxes. We will refer to these profiles as ``integrated'' because 
they are obtained by integrating over the whole extent of the  \ion{H}{ii} 
region. Also, in this figure, we marked with asterisks the
 WR star candidates (Armandroff \& Massey \cite{armandroff85}). Only one of 
these candidates was confirmed as a WR star; another was confirmed as the 
SNR Sandage 8 and the remaining were identified as OB stars (Azzopardi 
et al. \cite {azzopardi88}).

The integrated radial velocity profiles are, in general, single for all the 
\ion{H}{ii} regions; in this case, a single Gaussian function was fitted 
as described above. 
In only two cases: the SNR Sandage 8 (our source N18, see Table 
\ref{regiones}) and the giant shell Sandage 10/13 (N27), the 
integrated radial velocity profiles were complex. 
In those cases, we carried out a profile decomposition into two Gaussian
functions. 
We then list the two velocity components obtained from our profile fitting. 
A more detailed study of the kinematics of S8 from the FP data will be 
given elsewhere (Rosado et al. \cite{rosado00}). 
Also, a more detailed study of the internal kinematics of the superbubble 
N27 = 67 (HLG) and of other superbubbles in this galaxy is given in the 
following section.
  
Table \ref{regiones} gives a catalogue of the kinematical parameters of the 
\ion{H}{ii} regions of \object{IC 1613}, within the field of view of our 
observations. 
It is important to note that some of the objects listed in 
this table are ring--shaped and consequently are better classified as 
superbubbles and will be studied in more detail in the following section. 
We include these objects in Table \ref{regiones} because they appear in 
previous catalogues (Sandage \cite {sandage71} and HLG) but we exclude them 
from the calculation of the \ion{H}{ii} region luminosity function and diameter 
distributions.

In Table \ref{regiones}, we have listed the name (adopting both HLG's
and Sandage's conventions), diameter (see the definition for this parameter 
in Section 4.2), peak velocity and velocity dispersion from the Gaussian 
fitting to the radial velocity profile, and the H$\alpha$ flux, derived as 
described in Section 3.2.
To obtain luminosities, our fluxes were corrected only for galactic
foreground extinction using the $A_B$=0.21 mag value given in the RC2 catalog 
(de Vaucouleurs et al. \cite{devauco76}). 
The effects of internal extinction in these determinations seems to be not
crucial (Hippelein \cite{hippelein86}).
The radial velocities were corrected to the heliocentric system of reference.
Our dispersions ($\sigma_{obs}$) have been corrected for instrumental
($\sigma_{inst}$), thermal ($\sigma_{th}$) and intrinsic ($\sigma_{intrin}$)
 broadenings respectively, according to:

$$\sigma^2_{corr}= \sigma^2_{obs}-\sigma^2_{inst}-\sigma^2_{th}-
\sigma^2_{intrin}$$

\noindent where $\sigma_{inst}= 43.5$ km s$^{-1}$, 
$\sigma_{th}= 9.1$ km s$^{-1}$ for hydrogen gas at T$_e$=10$^4$
(Spitzer \cite{spitzer78}) and $\sigma_{intrin}=3$ km s$^{-1}$ for the fine 
structure of the H$\alpha$ line. Some authors also refer to $\sigma_{corr}$ 
as the turbulent velocity dispersion.
 
An inspection of this table shows that the heliocentric radial velocities range 
from --228 to --254 km s$^{-1}$.
The equivalent diameters range from 27 pc (corresponding to region N24) to
253.6 pc (for region  Sandage 3 or N9) and their mean
value is around 82 pc. 
The corrected velocity dispersions are supersonic for most of the objects
with values ranging from 16 to 35 km s$^{-1}$ with a mean value of 20 km s$^{-1}$.
Some \ion{H}{ii} regions have low S/N ratios and consequently their
determinations of peak velocity and velocity dispersion are uncertain. 
These cases have been quoted within parentheses in Table \ref{regiones} .
The H$\alpha$ fluxes range from 1.60$\times10^{-15}$ to 9.24$\times10^{-13}$
\mbox{ergs cm$^{-2}$ s$^{-1}$}; these lower and upper limits correspond to the
\ion{H}{ii} regions N13 and N27, respectively.

In Table \ref{stars} we give the stellar content in order to identify
the exciting star or stars that are related to the analyzed \ion{H}{ii} regions.
Column 1 gives the identification labels used in this work. Column 2 and 3
list the WR candidates (Armandroff \& Massey \cite{armandroff85}) that coincide
with some of our \ion{H}{ii} regions and their spectral type (or related object)
according to Azzopardi et al. (\cite{azzopardi88}).
Columns 4, 6 and 7 indicate the stellar associations and clusters determined
in previous works.
All but four \ion{H}{ii} regions have an evident stellar content that is probably
the source of the UV flux.

\subsection{The \ion{H}{ii} region luminosity function}

\begin{figure}[htb]
\includegraphics{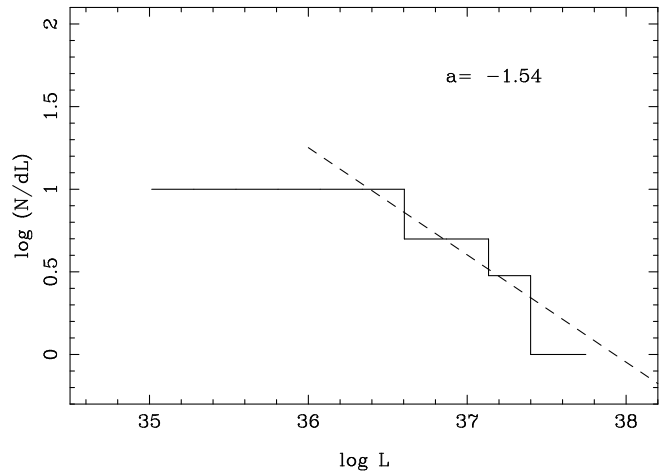}
\vspace{6.8cm}
\caption{ The H$\alpha$ luminosity function for the HII regions in \object
{IC 1613}. The dashed line is the least squares fit to a power law. The slope 
value is given to the right.}
\label{lum_function}
\end{figure}

Figure \ref{lum_function} shows the H$\alpha$ luminosity function for the 
\ion{H}{ii} regions in \object{IC 1613}.
For several galaxies, it has been shown that the brighter luminosity end
of the \ion{H}{ii} region luminosity function can be fitted with a power law in
the form: \mbox{$N(L)$=AL$^{a}$ $dL$ }, where $N$ is the number of \ion{H}{ii}
regions with luminosity $L$ in an interval of luminosity $dL$, and the slope 
$a$ is related to the morphological type of the galaxy (Kennicutt et al.
\cite{kennicutt89}), ranging from mean values of --2.3 (Sab-Sb galaxies) to 
--1.75 (Irregulars).
It also has been suggested that different ranges of the luminosity 
function are related to different kind of objects, such as \ion{H}{ii} regions
ionized by single stars and those ionized by clusters of stars (Hodge et al. 
\cite{hodge89}; Youngblood \& Hunter \cite{youngblood99}).

To obtain the \ion{H}{ii} region luminosity function we take the
logarithm of the number of \ion{H}{ii} regions divided by the corresponding
luminosity bin width (0.26 dex in our calculation). The bin width was chosen 
in terms of our relatively small number of objects (30 \ion{H}{ii} regions) and
the fit was calculated to the bins brighter than $\log L= 36.5$.
Our best fit slope for \object{IC 1613} gives $a=-1.54\pm 0.1$. 
This value is similar to the value ($a=-1.6\pm 0.2$) found by HLG, and is within
the mean values found for a sample of irregular galaxies studied by Kennicutt 
et al. (\cite{kennicutt89}), Strobel et al. (\cite{strobel91}), Kingsburgh \&
McCall (\cite{kingsburgh98}) and Youngblood \& Hunter (\cite{youngblood99}).

\subsection{The \ion{H}{ii} region diameter distribution}

We measured the diameter of the \ion{H}{ii} regions on the monochromatic
H$\alpha$ image and they are reported in column 4 in Table \ref{regiones}. 
The reported diameter is the equivalent diameter, which is defined as
$D=(4A/\pi)^{1/2}$. In this definition, $A$ is the area of the apertures 
wherein the photometry was performed. The measured diameters range
from 27 to 254 pc (this later value corresponds to large \ion{H}{ii} 
complexes that are not ring-shaped).

\begin{figure}[ht]
\includegraphics{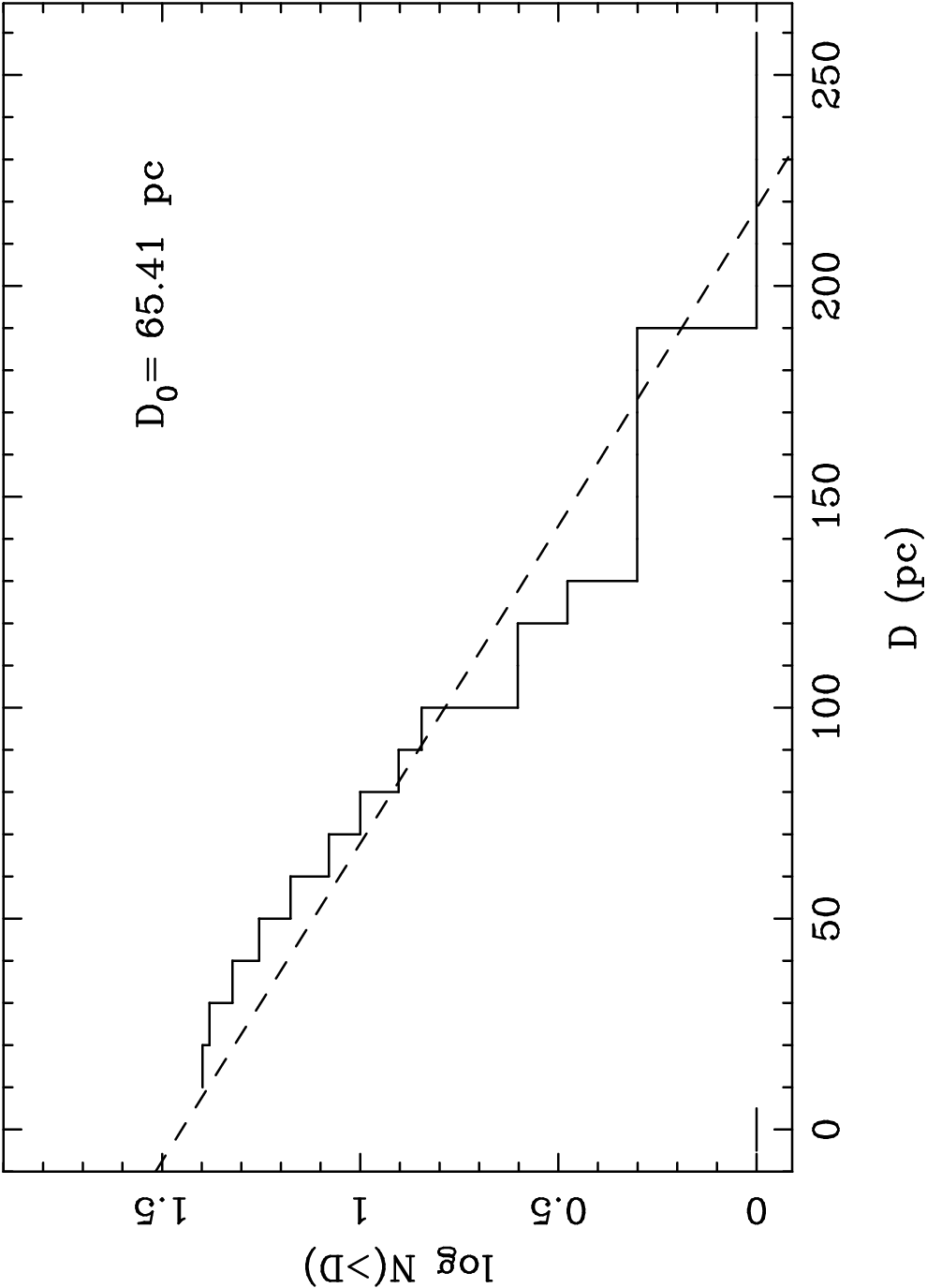}
\vspace{6.8cm}
\caption{The cumulative size distribution for the \ion{H}{ii} regions in
\object{IC 1613}, where the dashed straight line is a least squares fit. The
value for the characteristic diameter is given to the right.}
\label{distribution_HII}
\end{figure}

The cumulative \ion{H}{ii} region diameter distribution \hbox{$N(>D)$}  has been
calculated and in Figure \ref{distribution_HII} we plot \hbox{${\rm log}\,
N(>D)\, vs\, D$}.
This functional dependence was proposed by van den Bergh (\cite{vandenbergh81})
as a universal law for all galaxy types and it has been used to study the
statistics of the \ion{H}{ii} region populations in combination with the
\ion{H}{ii} region luminosity function. The slope in the cumulative
\ion{H}{ii} region diameter distribution is related to the characteristic
diameter $D_0$ of the \ion{H}{ii} regions in a galaxy.
For \object{IC 1613}, a least squares fit gives $D_0=65.41\pm 3$ pc.
This value is marginally in agreement with a previous one determined by HLG who 
obtained $D_0$=56 pc. However our value differs from that of Price et al.
(\cite{price90}) who quoted a larger value ($D_0$=83 pc). This discrepancy can 
be easily explained, because Price et al. (\cite{price90}) also considered
some superbubbles in their calculation (see Section 5).
Looking for other correlations, Hodge (\cite{hodge83}), using a sample of
18 irregular and related galaxies (\object{IC 1613} was not in his sample), found
a good correlation between the characteristic diameter $D_0$ and the absolute
magnitude $M_B$ of a galaxy.
In a subsequent work, Strobel et al. (\cite{strobel91}) increased the sample of
Hodge (\cite{hodge83}) with 7 additional galaxies (\object{IC 1613} was also not
included in these new objects). They found that the previous correlation between
$D_0$ and the absolute magnitude $M_B$ becomes weaker, having the form
\mbox{${\rm log}D_0=0.25(\pm 0.29)-0.098(\pm 0.017) M_B$}.
For \object{IC 1613} this relation gives a value of ${\rm log} D_0=1.64$ --
taking $M_B=-14.2$ from Mateo (\cite{mateo98}) -- in agreement with our
calculation (${\rm log} D_0=1.83$).
However, if we plot our resulting $D_0$ in Figure 11 in Strobel et al.
\cite{strobel91}, our data point falls outside the line fit in that plot, a fact
that points towards the conclusion given by Strobel et al. (\cite{strobel91}).  
In that sense, in a more recent work, Youngblood \& Hunter (\cite{youngblood99}),
using a sample of 29 irregular and Blue Compact Dwarfs (\object{IC 1613} not
included), have shown that this correlation is indeed a very weak one (see
their Figure 9).

\section{Kinematics of the superbubbles (giant shells) of \object{IC 1613}}

Figure \ref{obscura} shows one of the velocity maps (at V$_{helio}$=--234 km
s$^{-1}$) of the H$\alpha$ emission of \object{IC 1613}, obtained from our FP
observations.
The contrast of this image is higher than the direct image shown in Figure
\ref{direct} because the FP interferometer acts as a filter with a very narrow
width (0.41 \AA) while the bandwidth of the filter used to obtain the direct
image in Figure \ref{direct} is 20 \AA. 
Thus, it allows the detection of the line emission from the galaxy while reducing 
spurious line or continuum emission. 
Furthermore, in order to enhance the detection of low emissivity
features of large dimensions, this velocity map was smoothed spatially using 
a Gaussian filter of 4 pixels  width in both X and Y axis.

We note that  the entire observed field is covered by networks of shells, rings
and filaments whose diameters range from 110 pc (30\farcs6) to 325 pc (90\farcs6). 
We will follow the terminology of this type of structures given in the introduction. 
In accord with this terminology, Figure \ref{obscura} shows that the entire 
field of our observations of \object{IC 1613} is covered by superbubbles.
We were able to differentiate 17 large diameter ring-shaped structures.
The rough location of these superbubbles are marked on  Figure \ref{obscura}. 
We detect the seven superbubbles or giant shells, GS, (from GS1a and b to GS6) 
already reported by Meaburn et al. (\cite{meaburn88}), as well as several others 
located outside the field (to the S and to the W) covered by Meaburn et al.
(\cite{meaburn88}). Most of the newly detected superbubbles lie outside the
bright NE region of the galaxy and they are very faint and difficult to
visualize. Some of them are coincident with the \ion{H}{ii} regions and
filaments reported by Price et al. (\cite{price90}) and Hunter et al.
(\cite{hunter93}).

Figure \ref{ne-mapha} and  Figure \ref{ne-maps2} show some of the H$\alpha$ and 
[\ion{S}{ii}] velocity maps (or channels) where the network of superbubbles in 
the NE region is clearly appreciated. The heliocentric radial velocities for 
these superbubbles range from --100 to --350 km $s^{-1}$.
The superbubbles appear different in the several FP velocity maps, allowing us to
disentangle several of them that appear superimposed in the direct
images. 
In particular, the giant shells 1a and 1b of Meaburn et al. (\cite{meaburn88}) 
(corresponding to superbubbles R1 and R2, respectively) appear in different
velocity channels: R1 (Meaburn's GS 1a) is better visualized in channels
covering heliocentric velocities from --272 to --253 km s$^{-1}$ (and even more 
negative velocities not shown in the figures), whereas R2 (Meaburn's GS 1b) is 
better appreciated in channels where the heliocentric velocity ranges from 
--234 to --196 km s$^{-1}$.

In order to quantify the relative importance of the superbubbles and to compare
with other galaxies, we have calculated the 2-D porosity parameter, $Q_{2D}$, 
in our field, defined as the ratio of the total area occupied by the catalogued 
superbubbles to the total area of the galaxy seen in our field of view (Oey \&
Clarke \cite{oey97}). We obtain  $Q_{2D}$ = 0.75, a value that is intermediate
between the observed  $Q_{2D}$ values for the SMC (1.6) and M 31 (0.01) 
according with Oey \& Clarke (\cite{oey97}). However, the  $Q_{2D}$ of the whole
galaxy must await observations of the type described in this work covering the
whole dimension of \object{IC 1613}.

According to the procedure discussed at the end of Section 2, we obtained the 
radial velocity field of the superbubbles.
We used mainly the H$\alpha$ data; however, we used also the [\ion{S}{ii}] data
for the superbubbles which showed bright emission in the [\ion{S}{ii}] lines. 
Some of the superbubbles of the NE region appear also in Table \ref{regiones}. 
In that table we have only quoted the velocity values extracted from the 
integrated radial velocity profiles, while in the following we will use the 
radial velocity field in seeking for internal motions.

Figure \ref{profiles} shows, as an example, the radial velocity profile  of one
of the superbubbles where the splitting is clearly visible. 
As we can see from this figure, some of the profiles are complex and require 
to be fitted by a superposition of at least two Gaussian functions. 
We interpret the different components as being due to an expansion motion.  
Table \ref{cat_shells} gives a catalogue of the detected superbubbles, their
identification (if any) according to Meaburn et al. (\cite{meaburn88}), 
Price et al. (\cite{price90}) and Hunter et al. (\cite{hunter93}, hereafter
HHG), their angular dimensions, the peak heliocentric velocities of the 
Gaussian functions fitted to the velocity profile and the calibrated H$\alpha$
fluxes and luminosities.
The last column of Table \ref{cat_shells} corresponds to comments on the
morphology  of the superbubbles. We have found that some of the superbubbles
have the ``ring of \ion{H}{ii} regions'' structure (i.e., the shell has smaller
\ion{H}{ii} regions at its boundaries). This morphological type presents some
evidence of shock--induced star formation as Rosado et al. (\cite{rosado96}) 
have found for superbubbles in the LMC. 

\begin{figure*}[htb]
%\special{psfile=ms9979f6.ps angle=0 hscale=97 vscale=97 voffset=-578 
%hoffset=-25}
\vspace{13.9cm}
\caption{Velocity map (at V$_{helio}$=--234 km s$^{-1}$) of the  H$\alpha$ 
emission of \object{IC 1613}, obtained from the scanning FP observations. 
Superimposed on this map are the boundaries of the 17 superbubbles 
catalogued in Table \ref{cat_shells}.}
\label{obscura}
\end{figure*}

\begin{figure*}[htb]
\includegraphics{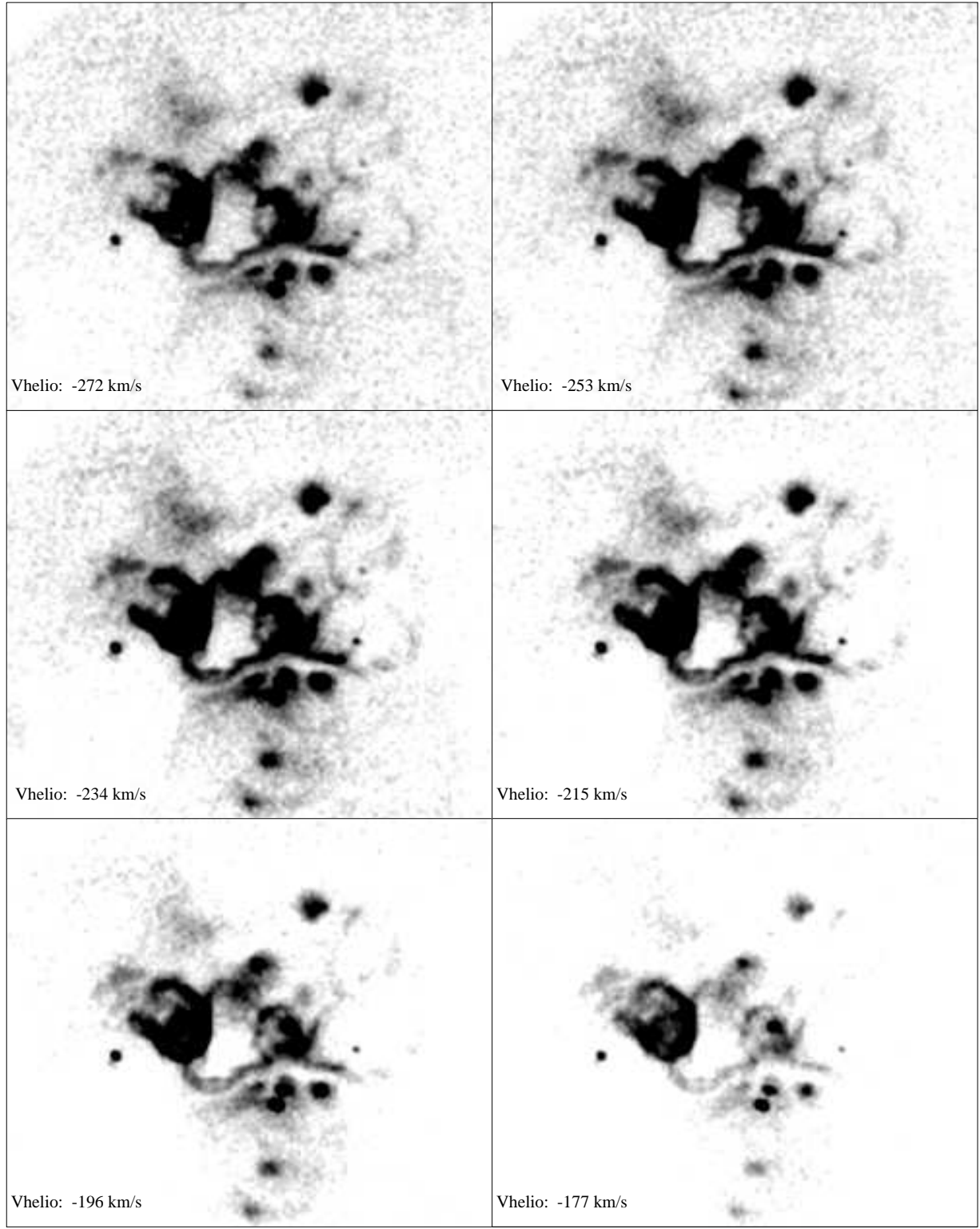}
\vspace{20.2cm}
\caption{H$\alpha$ velocity maps of the NE region in \object{IC
1613} where the superbubble network appears bright. In this map, the 
superbubbles R1, R2, R4, R5, R6, R7, R8, R9, R10 and R13 are more clearly
visible because of the high contrast of the print.}
\label{ne-mapha}
\end{figure*}

\begin{figure*}[htb]
\includegraphics{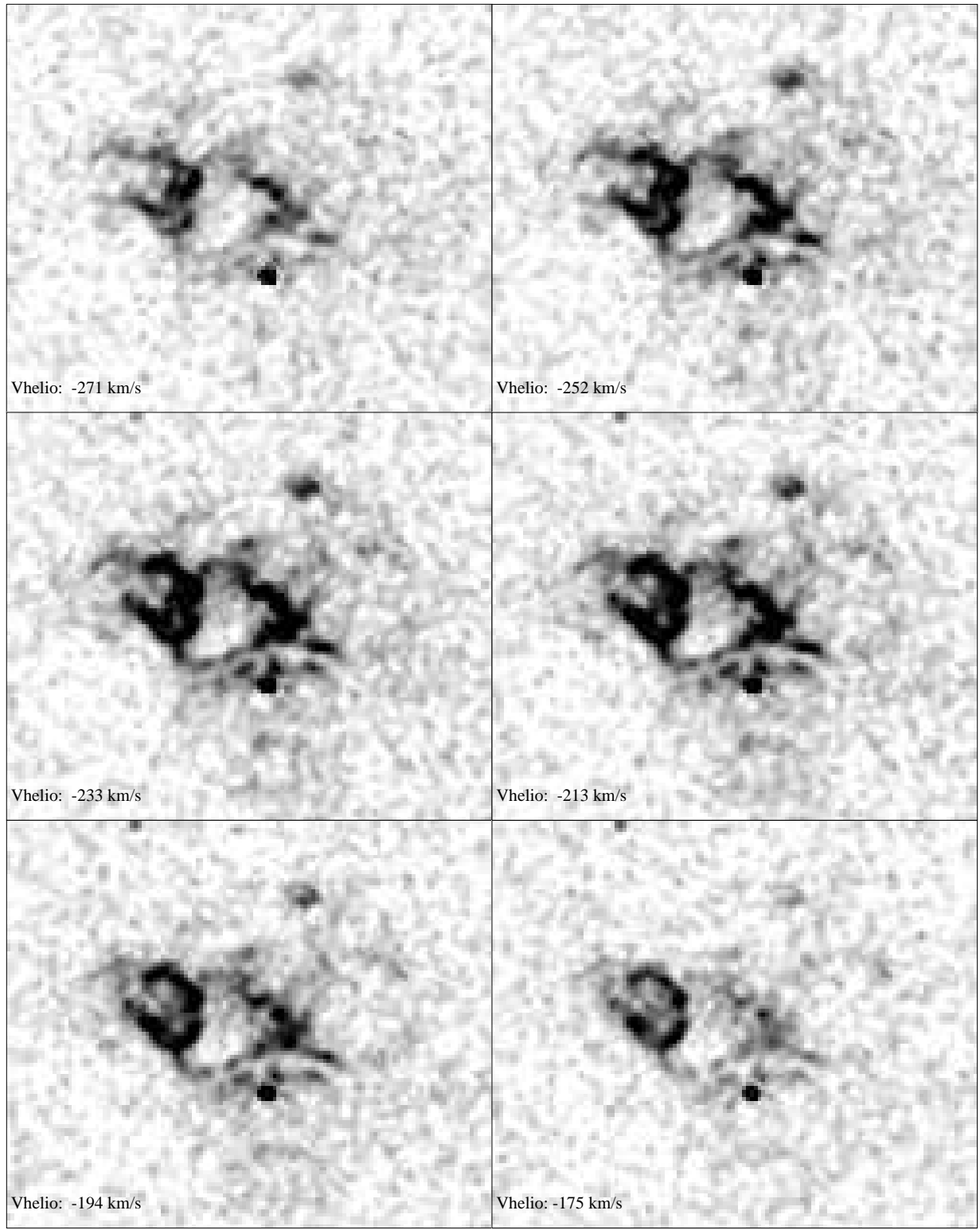}
\vspace{20.2cm}
\caption{ [SII] velocity maps of the NE region in \object{IC
1613} corresponding to the same velocities as in Figure \ref{ne-mapha}. 
The bright feature to the South corresponds to the SNR Sandage 8.}
\label{ne-maps2}
\end{figure*}

\subsection{ The stellar content of the superbubbles}

The superbubbles catalogued in Table \ref{cat_shells} are directly related to
the stellar associations of \object{IC 1613}. These stellar associations were
catalogued by Hodge (\cite{hodge78}) for the entire galaxy. In Paper I, we have
studied the stellar content of the NE region of the galaxy and applied an
automatic method to obtain the boundaries of the stellar associations.
Figure \ref{stars_and_shells} shows the stellar associations studied in 
Paper I, where the superbubble boundaries shown in Figure \ref{obscura} are 
also marked. 
We see that there is a good agreement between the locations of most of the
stellar associations and the centers of the superbubbles. For the superbubbles
outside the NE region we also find  a good agreement between the positions of
Hodge's (\cite{hodge78}) associations and the superbubbles.

The stellar associations within the superbubbles boundaries are reported
in Table \ref{stars_shell}. In this table we have listed (1) the superbubble's
number (according to our catalogue), (2) the Paper I and Hodge's (\cite{hodge78}) 
stellar associations located within the superbubble boundaries, (3) the number 
of blue stars of the interior associations (stars with $M_{v}$ $<$ --2.0, 
$(B-V)_{0}$ $<$ 0.0 and $(U-B)_{0}$ $<$ --0.5, corresponding to stars with 
masses larger than 7  M$_{\odot}$ , (4) the age of the stellar associations, the
possible existence of (5) OB stars and (6) blue supergiants with known spectral
types and (7) the $\Gamma$ exponent of the Initial Mass Function of the interior
associations.
Columns 3, 4, 6 and 7 were derived from the data published in Paper I. Column 5
is based on the spectra published for several Wolf-Rayet (WR) star candidates in
this galaxy (Armandroff \& Massey \cite{armandroff85}, Azzopardi et al. 
\cite{azzopardi88}). An examination of this table shows that the stellar 
associations reported in Paper I have a better correlation  with the catalogued 
superbubbles than Hodge's.
This is because Hodge's  (\cite{hodge78}) associations are, in general, larger 
than those reported in Paper I. Given that so many of them are interior to the 
superbubbles, we can conclude that the method of identification of stellar 
associations discussed in Paper I works very well. 
We also see that every catalogued superbubble  can be associated with a 
relatively young stellar association (except R4, which probably has an
older interior association), even the dimmer superbubbles, located far away from 
the bright NE region of \object{IC 1613}. This suggests that the interior 
associations are responsible for the formation of the superbubbles.

\subsection {Physical parameters of the superbubbles}

From the observed data (listed in Table 6) we can extract several physical
parameters of the superbubbles:

1. The rms electron density of a superbubble is calculated from its H$\alpha$
flux and dimensions according to the relation: 

$$n_{e}^{2}({\rm rms)\ (cm}^{-6})=2.74 \times 10^{18}\ {\rm F}({\rm H}\alpha)\ 
{\rm R}^{-1}\ \theta^{-2}$$

This relation was obtained by assuming that the emitting gas is at
T = 10$^{4}$ K and distributed in an spherical shell of radius R (in pc) and
thickness, $\Delta$R = R/12. The other quantities, F(H$\alpha$) and
 $\theta^{2}$,
are listed in Table \ref{cat_shells}.

2. The expansion velocity, V$_{EXP}$, of a superbubble can be estimated as one
half of the difference in velocities of the components listed in Table
\ref{cat_shells}. 

3. The mechanical energy of the superbubble, K.E. = 0.5 M  V$_{EXP}^{2}$, is
calculated according with the relation:

$${\rm K.E. (ergs)}=1.7 \times 10^{41}\ {\rm n}_{e}({\rm rms)\ ( cm}^{-3})\ 
{\rm V}_{EXP}^2({\rm km\ s}^{-1})\ {\rm R}^{3}$$

In obtaining this relation we have assumed that the mass is concentrated in the
spherical shell of radius R (in pc) and thickness, $\Delta$R = R/12. 
We have also assumed that $\mu$ = 0.65, as for an ionized hydrogen gas.

4. The kinematic age of the superbubble can be obtained from the following
relation:

$$ {\rm t (yr)}=4.8 \times 10^{5}\ {\rm R}/{\rm V}_{EXP}\ ({\rm km\ s}^{-1})$$

This relation has been obtained by noting that the kinematic age of wind blown
bubbles can be estimated as: t = 0.5 - 0.6 R/V (McCray \cite{mccray77}).

Table \ref{parameters_shells} lists the physical parameters of the catalogued
superbubbles calculated as discussed above: the superbubble's number
(according to our catalogue), the shell radius (see Sect. 5.3), the 
rms electron density, the expansion velocity, the mechanical energy 
and the dynamical age of the superbubble.
Meaburn et al. (\cite{meaburn88}) have obtained expansion velocities for five of
these superbubbles. Our results agree with these previous observations except for
the dimensions of some of the superbubbles.

\subsection{The diameter distribution of the superbubbles}

The angular dimensions of the superbubbles were determined on the velocity
maps at H$\alpha$ and they are given in column 3 in Table \ref{cat_shells}.
The reported dimensions in this table correspond to the extension of the
semiaxes in arc sec. In the cases where we have two different values, i.e., when 
the shell is not circular, we average them. The resulting linear radii are listed 
in column 2 in Table \ref{parameters_shells}.

Comparing the size distribution of the \ion{H}{ii} regions and \ion{H}{ii}
complexes of \object{IC 1613} (see Section 4.2) against the superbubble size
distribution we note the following differences: while the diameters of the
\ion{H}{ii} regions range from 27 to 254 pc (this later value corresponds to
large \ion{H}{ii} complexes not ring-shaped), those of the superbubbles range
from 110 to 304 pc (note that there is an overlap between these two populations).
Thus, the superbubbles have, in general, larger dimensions than those of
\ion{H}{ii} regions, pointing towards a more evolved nebular population.
Figure \ref{histogram} shows the size distribution, $N(R)-R$, of 
the superbubbles in \object{IC 1613}. 
It is worth comparing this size distribution with the size distributions of 
other galaxies. 
In order to do so, we have examined the size distributions of the galaxies 
SMC, Holmberg II, M31 and M33 reported in Oey \& Clarke (\cite{oey97}; see their 
Figures 3 and 4).
It is important to take into account the small number statistics of the
superbubbles in the studied field of \object{IC 1613} (only 17). Furthermore, our
superbubbles are ionized regions while the superbubbles in Oey \& Clarke
(\cite{oey97}) correspond to \ion{H}{i} holes. 
Nevertheless, we see that the distributions show some similarity. The peak of 
maximum frequency of the superbubbles in these galaxies corresponds to 
$D\approx 80$ pc (SMC), 120 pc (M33), 180 pc (M31) and 500 pc (Hol II); this peak, 
in the case of \object{IC 1613}, is at $\approx 100 - 140 $ pc, thus, similar 
to the one of M 33 \ion{H}{i} holes. 
On the other hand, there is a difference in the maximum linear extent of the 
superbubbles: the superbubbles in \object{IC 1613} reach $D\approx  300$ pc while 
the superbubbles in the other studied galaxies reach diameters of 720 pc (M31), 
760 pc (SMC), 1160 pc (M33) and 1400 pc (Hol II); note that the latter two values 
fall into the category of supershells according to the typical terminology. 
This difference could be due to the fact that we are only detecting ionized shells 
while the other studies are based on \ion{H}{i} holes. 
We have also obtained the cumulative diameter distribution $N(>D)$ for the
superbubbles. In Figure \ref{distribution_bubbles} we plot ${\rm log}\,
N(>D)\, vs\, D$. For the superbubbles of \object{IC 1613} a least squares fit
gives $D_0=91.84\pm 3$ pc. This value is in agreement with the value
($D_0$=83 pc) determined  by Price et al. (\cite{price90}) who included some
superbubbles in their calculation of the diameter distribution. 

 \begin{figure}[ht]
\includegraphics{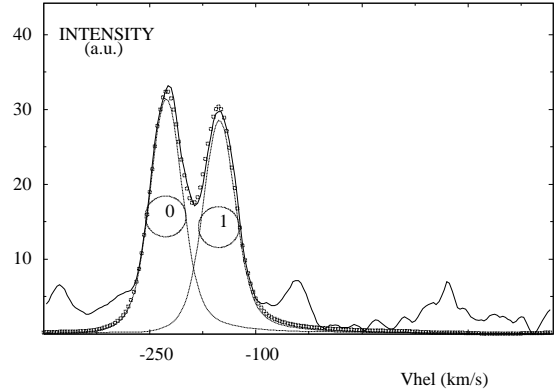}
\vspace{5.8cm}
\caption{Radial velocity profile of one region of superbubble R2 catalogued in 
Table \ref{cat_shells} and its decomposition in two Gaussian functions, (0) 
and (1) with peak radial velocities of \hbox{-- 244} and  \hbox{-- 147} 
\hbox{km s$^{-1}$}, respectively, for this particular case. 
The horizontal axis plots the heliocentric radial velocity while the vertical
axis corresponds to the intensity (in arbitrary units). 
The low velocity peaks correspond to night-sky lines mixed with diffuse
emission.}
\label{profiles}
\end{figure}

\begin{figure*}[ht]
\includegraphics{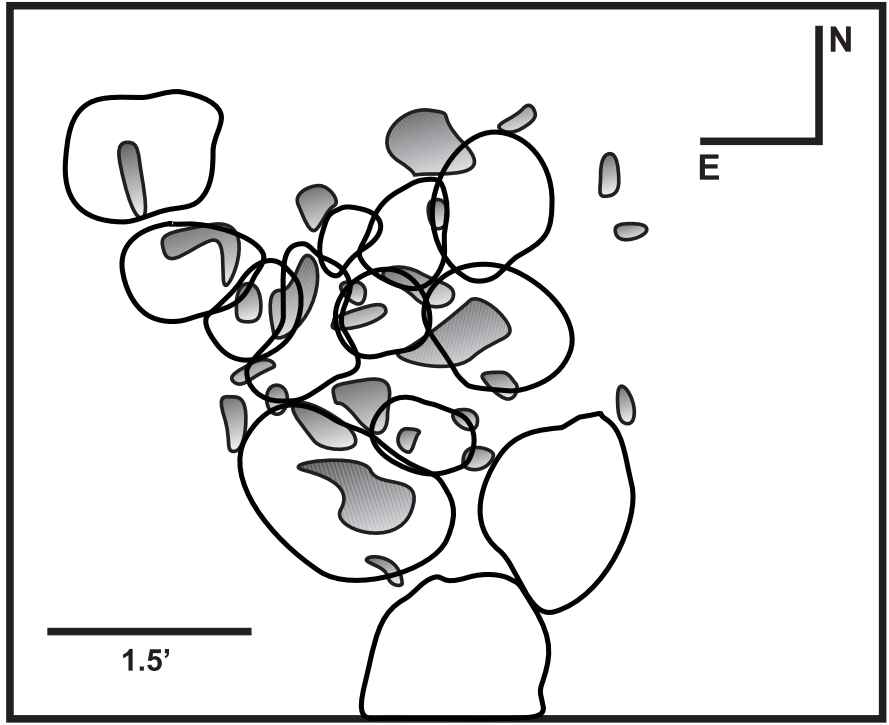}
\vspace{10.0cm}
\caption{Sketch  of the stellar associations studied in Paper I (shadowed
regions). We have superimposed on this some of the superbubble boundaries
already shown in Figure \ref{obscura}. The scale and the orientation are
indicated down to the left and top to the right, respectively.}
\label{stars_and_shells}
\end{figure*}

\subsection {Discussion}

We have seen in the preceding sections (more precisely in Section 5.1) that 
every superbubble detected in this work corresponds to one (or several) stellar
association(s) interior to the superbubble. This suggests that the superbubbles
are formed by the combined action of stellar winds and supernova explosions of
the massive stars of the associations. In what follows, we will examine in more
detail this possibility and the models used to describe this formation and
evolution.

The shell formation and evolution as a result of the interaction of strong winds
of a massive star with its ambient medium has been studied a long time ago by
several authors (Dyson \& de Vries \cite{dyson72}; Castor et al. \cite{castor75};
Steigman, Strittmatter \& Williams  \cite{steigman75}; Weaver et al. \cite{weaver77}, 
amongst others).
The more favored (and complete) description of this interaction is the 
``standard model'' (Weaver et al. \cite{weaver77}) which considers that a star
at rest releases a {\it constant} wind power, L$_{W}$, that interacts
with the ambient medium (assumed to be homogeneous).  
As a result of this interaction, two shocks are formed (one in the stellar wind
and another in the ambient medium). The shell formation is the result of these
shocks and the shell expansion is driven by the thermal pressure of the shocked
stellar wind.
The shocked stellar wind has no significant radiative cooling and for this
reason this model is also called {\it adiabatic} or {\it energy conserving}.
This model could also be applied to explain the formation of larger shells
(superbubbles and supershells) formed by the combined action of stellar winds and
supernova explosions. In spite of the popularity and theoretical simplicity of
the standard model, several observational difficulties have prevented its test in
real cases. The main difficulty is the poor knowledge of the interior stars and
of their wind power. 
Observational works (Oey \cite{oey96b}) where the stellar content has been 
thoughtfully analyzed and using numerical (instead of the analytical) models 
that take into account the temporal variations of the wind power as a result of 
the stellar evolution  (Oey \cite{oey96a}), have found discrepancies between
the predicted shell radius and expansion velocity in the sense that the expansion
velocities are too high for the observed radii (``dynamical discrepancy'') or
between the shell radius and velocity and the wind power, L$_{W}$, and pre-shock
density, n$_{0}$, in the sense that L$_{W}$/n$_{0}$ seems to be overestimated by
a factor of 10 (``growth rate discrepancy''). 
Having these considerations in mind we will proceed to analyze our results, 
summarized in Table \ref{parameters_shells}.

A close examination of the physical parameters reported in Table
\ref{parameters_shells} shows that:

-- The rms electron densities of the superbubbles are quite low (from 0.3 to up
to 6.3 cm$^{-3}$). This makes it quite difficult to obtain electron densities from
[\ion{S}{ii}] line-ratio diagnosis because the line-ratios will fall in the low
density regime (n$_{e}$ $<$ 300  cm$^{-3}$).

-- The H$\alpha$ luminosities of the superbubbles vary from 2.11 $\times$ 10$^{36}$ 
to 5.89 $\times$ 10$^{37}$ erg s$^{-1}$. These luminosity values imply a number 
of ionizing photons s$^{-1}$ ranging from \hbox{1.55  $\times$ 10$^{48}$} to 4.33  
$\times$ 10$^{49}$. These ionizing photons can be easily supplied by the interior 
blue stars. Thus, the superbubbles seem to be ionized by the stars of the interior 
associations.

-- The dynamical ages of the superbubbles (column 6 of Table
\ref{parameters_shells}) are of the order of 1-2 Myr and in every case, shorter
than the age of the interior stellar association (column 4 of Table \ref{stars}).
The fact that the dynamical ages are shorter than those of the stellar
associations suggests that the stellar associations could form the superbubbles.
Furthermore, the fact that there is a delay (of about 10 Myr) between the
formation of the stellar association and the formation of the superbubble
(estimated by its dynamical age), suggests that only the most massive stars of
the interior association are responsible for superbubble formation when they
evolve into supergiants (or perhaps have exploded as supernovae). 

-- The obtained expansion velocities of the superbubbles are in general quite
high. All the known expansion velocities are larger than 25 km s$^{-1}$ and there
are two superbubbles with expansion velocities larger than 40 km s$^{-1}$.
This is similar to the case of the superbubbles in the LMC where Rosado et al.
(\cite{rosado81}; \cite{rosado82}), Rosado (\cite{rosado86}) and Oey (\cite{oey96a}) 
have found several superbubbles with high expansion velocities.
 The kinematics of this class of superbubbles (``high velocity superbubbles'')
cannot be explained in terms of the ``standard model'' of adiabatic wind-blown
bubbles (Weaver et al. \cite{weaver77}) because the expansion velocities are too
large for the observed shell radius, according to the predictions of this
model.

\begin{figure}[ht]
\includegraphics{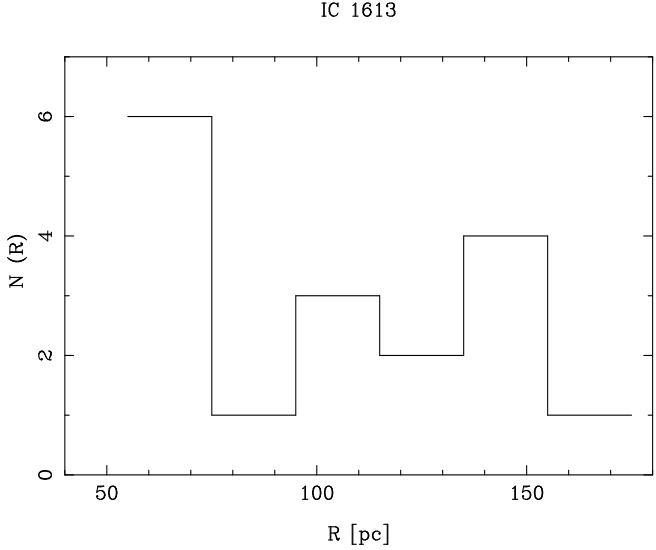}
\vspace{7.5cm}
\caption{Size distribution function,  $N(R)-R$, for the superbubbles in 
\object{IC 1613}. }
\label{histogram}
\end{figure}

\begin{figure}[ht]
\includegraphics{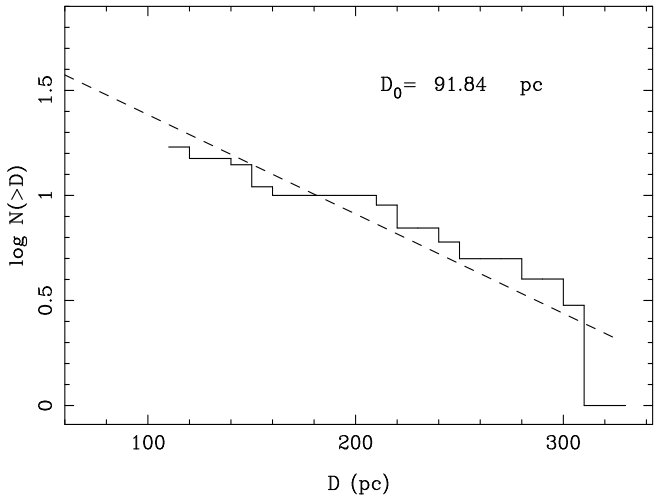}
\vspace{7.5cm}
\caption{The cumulative size distribution for the superbubbles in 
\object{IC 1613}. The dashed straight line is a least square fit. The value 
for the characteristic diameter is given to the right.}
\label{distribution_bubbles}
\end{figure}

In the case of the superbubbles of \object{IC 1613}, our stellar photometry can
give us some knowledge of the massive stars interior to the catalogued
superbubbles but we cannot follow the treatment given in Oey (\cite{oey96a})
because further spectroscopic work is needed in order to know accurately the
content of massive stars. 
In the absense of this accurate knowledge we can, nevertheless, estimate the 
required wind power (L$_{W}$) of each superbubble and compare it with the 
number of massive stars that we detect in our photometric observations, in 
order to see, in terms of energy considerations, whether the standard model 
could explain the superbubble formation.

Assuming that the superbubbles evolve according to the standard model of
wind-blown bubbles (Weaver et al. \cite{weaver77}), the wind power, L$_{W}$,
released by the stars {\it constantly during all} of the dynamical age of the shell
is given by:

$${\rm L}_{\rm W} = 3.2 \times 10^{-7} {\rm n}_0 ({\rm cm}^{-3}) {\rm V_{EXP}}^3
({\rm km\ s}^{-1})\ {\rm R}^2\times 10^{36}\ {\rm ergs\ s}^{-1}$$ 

\noindent where,  n$_{0}$, the ambient pre-shock density, is taken as:  
\mbox{n$_{0}$ =  n$_{e}$ / 4} by assuming  that the swept-up mass in the shell
comes from a homogeneous sphere of radius R.  Both n$_{e}$ and R are listed in
Table \ref{parameters_shells}. 

A wind power of 4$\times$10$^{36}$ ergs s$^{-1}$ is reasonably expected for one
O9.5 or B0 supergiant star (Pauldrach et al. \cite{pauldrach86}) during its
lifetime in this phase (about  10$^{6}$ yr). More massive stars can have larger
wind powers but the duration of those phases is ten times shorter. That is the
case for O4I stars that can release more than  100 times the wind power of
1$\times$10$^{36}$ ergs s$^{-1}$, or of WR stars that can release about
50$\times$10$^{36}$ ergs s$^{-1}$; however, these events occur only during time
scales shorter than the dynamical ages of the superbubbles considered here
(a few 10$^{5}$ yr compared with the dynamical ages of about 1 Myr derived in this
work).

In Table \ref{wind_power} we list the required  wind power L$_{W}$ that can
explain the observed expansion velocities and radii of the superbubbles studied
here, in terms of the standard model, and  the number of detected blue stars (with
M$_{V}$ $<$ --2, or M$_{*}$ $>$ 7 M$_{\odot}$), N$_{B}$, quoted in column 3 of
Table \ref{stars_shell}. 
We see that superbubbles with expansion velocities lower than 30 km s$^{-1}$ have
wind powers that are easily explained by the number of blue stars inside them
(assuming that they are all supergiants and earlier than  B0). On the
contrary, there are some extreme cases, such as R2, where the required wind power
seems to imply that some of the interior blue stars should release much more 
powerful winds. 
Given our limitations in the accurate knowledge of the stellar content of
the superbubbles, we can only state that, from energy considerations, the
superbubbles studied here could be formed by the stellar winds of the interior
stars. 

It is interesting to note from Table \ref{cat_shells} that the superbubbles of
\object{IC 1613} with higher velocities seem to be brighter in [\ion{S}{ii}]
emission. On the other
hand, an inspection of Table \ref{parameters_shells} shows that all the
superbubbles have mechanical energies of at most 80 $\%$ the energy of one
supernova explosion (approximately 1--3$\times$10$^{51}$ ergs). 
Taking into account that between 20 to 30 $\%$ of this energy is transmitted to
the ambient medium as kinetic energy, one or two SN explosions could also account
for the formation (or acceleration) of the superbubbles. 
However, no traces of these SN explosions (such as extended X-rays sources or
non-thermal radio emission) have been found.

\section{The velocity field of \object{IC 1613}} 

\begin{figure*}[htb]
\includegraphics{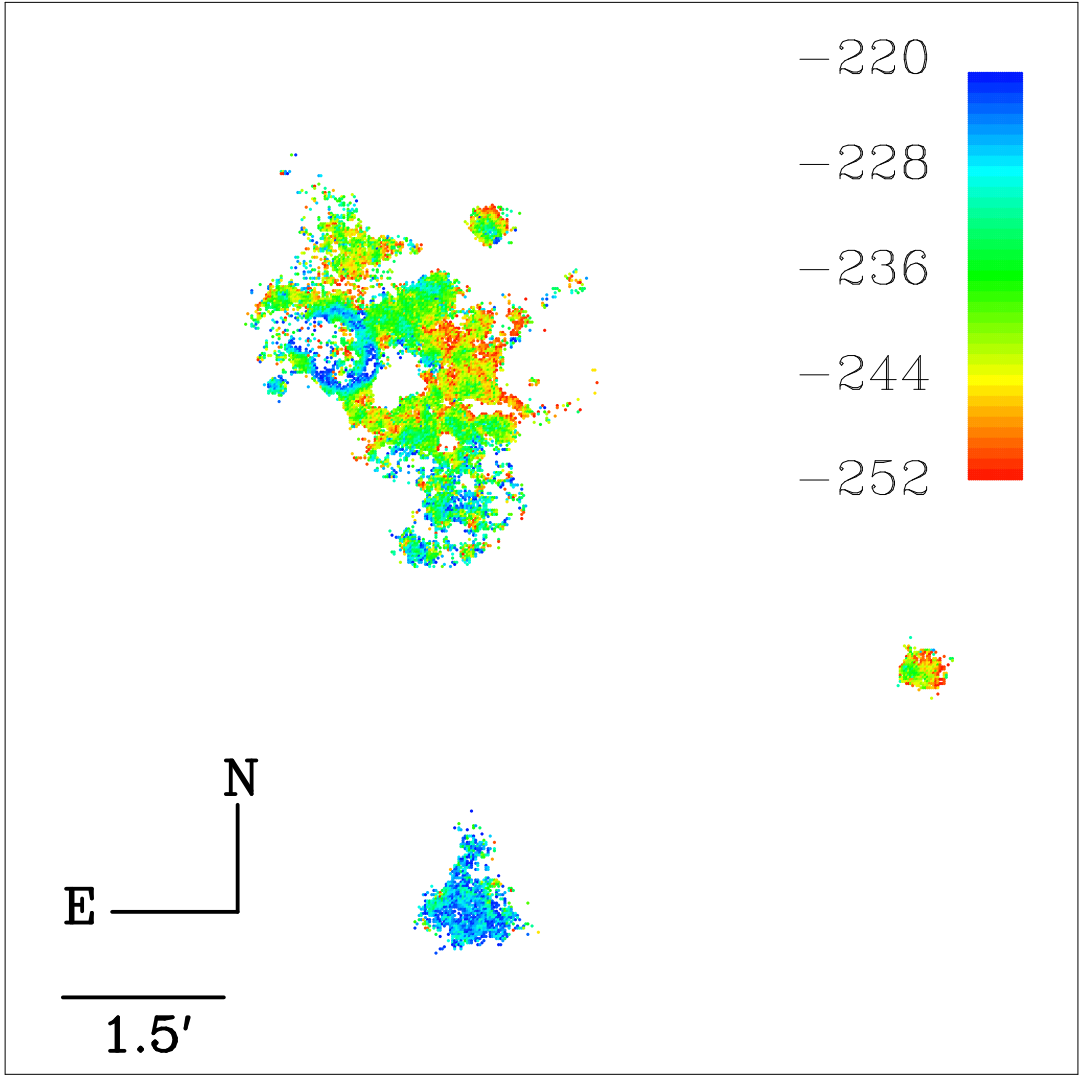}
\vspace{15.0cm}
\caption{\object{IC 1613} H$\alpha$ observed radial velocity distribution (see
text for details). The colors are related to velocities given in 
\mbox{km s$^{-1}$}. The scale and the orientation are indicated to the
left.}
\label{vr}
\end{figure*}

Figure \ref{vr} shows the global radial velocity field (not corrected for 
projection effects) of the central region of \object{IC 1613} obtained 
from our FP data in H$\alpha$, considering diffuse gas, \ion{H}{ii} 
regions and superbubbles.
In this map a chaotic distribution of velocities across the galaxy can be 
seen: some local events are superimposed onto the regular field. These local
events are the peculiar kinematical signatures of the superbubbles, which 
are notorious. 
However, when the expansions are extracted, we see a variation in the mean
velocities across the field.
Along the NE region, a gradient of velocities ranges from --210 km s$^{-1}$ in 
the southeast to --250 km s$^{-1}$ in the northwest.
The extreme detected velocities are due to the 
contamination of the superbubble and SNR internal motions. 
Indeed, the region N27 (with the maximum velocity of $-$139 km s$^{-1}$) is 
the superbubble R2 and N18 (with the minimum velocity of $-$328 km s$^{-1}$) is 
the SNR Sandage 8. It would be important to have kinematic data of the ionized 
gas over the whole dimensions of this galaxy in order to obtain its rotation 
curve.

\begin{figure}[ht]
\includegraphics{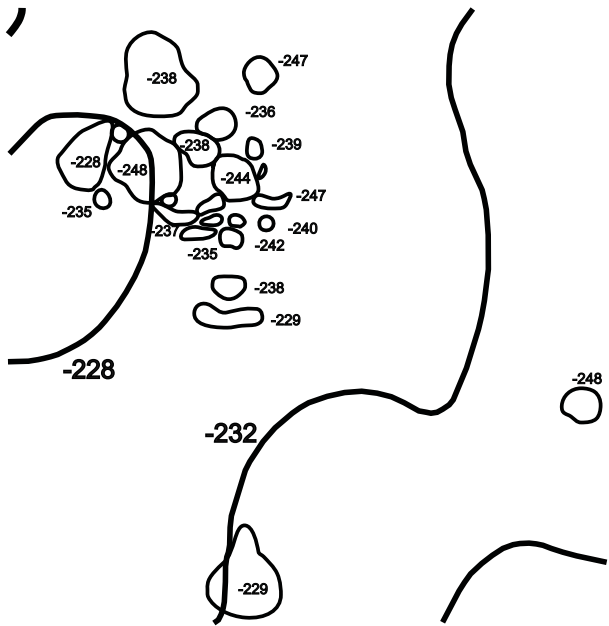}
\vspace{9.0cm}
\caption{\object{IC 1613} H$\alpha$ observed radial velocity distribution. 
Heavy lines correspond to \ion{H}{i} isovelocities (Lake \& Skillman
\cite{lake89}).}
\label{vel_HI}
\end{figure}

On the other hand, in Figure \ref{vel_HI}, the 21-cm isovelocities from 
Lake \& Skillman (\cite{lake89}) are sketched together with the radial velocity 
distribution of the \ion{H}{ii} regions, obtained in this work.
Their velocity resolution is 6.18 km s$^{-1}$ and the linear resolution is 
approximately 210 pc. In this figure we can see that in general, the optical 
distribution does not correspond to any particular kinematical \ion{H}{i} feature. 
However, with respect to the density distribution (not shown), it has been noted
that the optical extent of \object{IC 1613} is located at the peak of the
\ion{H}{i} density distribution and at the ridge of an \ion{H}{i} hole 
(Lake \& Skillman \cite{lake89}). In order to compare the kinematics of the 
ionized gas with the  \ion{H}{i} kinematics we also require more observations 
covering the whole optical extent of this galaxy.

\section{Conclusions}

We have measured -- by means of a Fabry-Perot interferometer -- the kinematical 
parameters of the \ion{H}{ii} regions that we have detected within the 
10$\arcmin$ field of view of our observations (which includes most of the 
optical emission of \object{IC 1613}). 
We found that the \ion{H}{ii} regions of \object{IC 1613} present similar 
integrated properties as well as H$\alpha$ line profiles with simple 
structure that are fitted by single Gaussian curves. 
We obtained also the H$\alpha$ fluxes and luminosities of these 
\ion{H}{ii} regions. The luminosity function of these  \ion{H}{ii} regions can be
fitted to a power law function of slope $a=-1.54\pm 0.1$ in agreement with the 
value given by HLG. We have also studied the diameter distribution and the 
cumulative diameter distribution of the \ion{H}{ii} regions within our field of
 view. We obtained a characteristic diameter $D_0=65.41\pm 3$ pc which is 
slightly larger than the previous determination by HLG, probably because our 
field is smaller. Our characteristic diameter value does not seem to follow 
the correlation found by Hodge (\cite{hodge83}) between $D_0$ and the absolute 
magnitude $M_B$ for a sample of irregular galaxies.

We have also found that the ISM of the entire galaxy is organized in a complex
network of superbubbles. 
 We have catalogued the superbubbles detected in our
field of view. We have also calculated the 2-D porosity parameter,  $Q_{2D}$, in 
the field of view of our observations and seen that the $Q_{2D}$ value is 
intermediate between the observed values for the SMC and M31.

Most of the superbubbles that we have detected show complex radial velocity
profiles (some others are so faint that we cannot measure line-splitting in their
velocity profiles). These kind of profiles along the same nebula are indicative
of internal gas motions and expansions. We have determined the expansion
velocities and other physical parameters of the superbubbles found in this work.
 In addition, we have used the photometric data published in a previous work
(Paper I) and in Hodge (\cite{hodge78}) in order to know the stellar content of the
catalogued superbubbles.
We find that almost every superbubble has an interior association containing
massive stars, suggesting a physical link between them. Furthermore, we find that
the dynamical time scales of the superbubbles are always shorter than the ages 
of the interior stellar associations.  The mean temporal delay between those ages
(about 10 Myr) suggests that only evolved members (i.e., the more massive stars)
of these associations are responsible for the formation of the superbubbles,
possibly when they evolve to supergiants or when they explode as supernovae.

We have found that all the superbubbles with measured expansion velocities have
expansion velocities larger than 25 km s$^{-1}$. We used energy considerations 
in order to explain the origin of the superbubbles in terms of stellar winds and
supernova explosions.
We find that the superbubbles with extreme expansion velocities (larger than
36 km s$^{-1}$) require interior stars with winds much more powerful
than the average of B0--O9.5 supergiant stars if they are wind blown shells 
following the ``standard model''.  On the other hand, one or two supernova 
explosions could explain the mechanical energies involved. However, no other 
traces of SN explosions (extended X-rays, non-thermal radio emission) are 
known for these superbubbles except an enhancement of their [\ion{S}{ii}] 
emission.

We have obtained the size distribution and the cumulative diameter function of
superbubbles detected in our field of view and compared these size distributions
with size distributions of superbubbles in other galaxies. Our results are 
statistically poor. However, the value of the characteristic diameter we obtain,
$D_0=91.84\pm 3$ pc, is in agreement with the value given by Price et al.
 (\cite{price90}) for the nebulae in this galaxy.

On the other hand, we have determined the global H$\alpha$ velocity field of
\object{IC 1613}, inside our 10$\arcmin$  field of view. The velocity field is 
contaminated by the internal motions of the superbubbles of this galaxy. 
Nevertheless, a gradient of velocities is suggested across the SE-NW direction.
In addition, we compare the mean velocities obtained from the  \ion{H}{ii} 
regions with the  kinematics of the \ion{H}{i} gas but we do not find a clear 
agreement between them. 

\acknowledgements{

The authors wish to thank Drs. I. Puerari, D. Mayya, A. Luna
and O. L\'opez--Cruz, for their comments and criticism about this work, 
and Mrs. J. Benda for the revision of the final version. 
MVG thanks support from a CONACyT scholarship under register number
114735. 
This work was partially supported by the grants IN122298 of DGAPA-UNAM 
and 27984-E of CONACYT. 

This research used the NASA/IPAC Extragalactic Database (NED) which
is operated by the Jet Propulsion Laboratory, California Institute of
Technology, under contract with the National Aeronautics and Space
Administration.}

\begin{table*}
\caption []{ IC 1613 properties}
\label{properties}
\begin{tabular}{ll}
\hline
 Name  			                               & IC 1613    \\
 Type$^{\mathrm{a}}$    	                       & IB(s)m     \\
 R.A (1950)$^{\mathrm{a}}$                             & 01h02m11.8s \\
 Dec (1950)                                            & +01d51m00s  \\
 B magnitude$^{\mathrm{a}}$                            & 9.88        \\
 Angular size (arcmin)$^{\mathrm{a}}$               & 16.2$\times$ 14.5 \\ 
 Heliocentric radial velocity (km s$^{-1})$ $^{\mathrm{b}}$ & -237         \\
 Distance (kpc)$^{\mathrm{c}}$                               & 725          \\
 Inclination (degrees)$^{\mathrm{b}}$                  & 30           \\
 P.A  (degrees)$^{\mathrm{b}}$                         & 50         \\
\hline
\end{tabular}
\begin{list}{}{}
\item[$^{\mathrm{a}}$] NED
\item[$^{\mathrm{b}}$] Optical heliocentric radial velocity from Mart\'\i n (\cite{martin98}) 
\item[$^{\mathrm{c}}$] Adopted distance in this work, based on Freedman
(\cite{freedman88a}, \cite{freedman88b}) 
\end{list}
\end{table*}

\begin{table*}
\caption []{ Parameters of the FP Interferometric Observations}
\label{observations}
\begin{tabular}{lll}
\hline

 Instrument  			& PUMA     	   	&	\\
 Detector    			& TEKTRONIX CCD   	&       \\
 FP Scanning Steps       	& 48               	&       \\
 Finesse     			& 24               	&        \\
\hline
\hline
%\vspace{.4cm}
 Filters               	&  H$\alpha$ 	& $[\ion{S}{ii}]$   	\\
 Central Lambda (\AA)   &  6570 		& 6720	\\ 
 Interference order   	& 330 			& 322 		\\
 Free spectral range (km s$^{-1}$) 		& 847   & 931   \\
 Sampling resolution  (km s$^{-1}$/channel)	& 18.9	& 19.4  \\
 Calibration line       & 6562.78 H 		&  6717.04 Ne  	\\
 Binning              	& 2$\times$2  		& 4$\times$4  	\\
 Final pixel size (arcsec)  & 1.17			& 2.34	\\
\hline

\end{tabular}
\end{table*}

\begin{table*}
\caption[ ]{HLG HII regions taken for calibration}
\label{standard}
\begin{tabular}{llllll}
\hline
Sandage$^{\mathrm{a}}$ & HLG$^{\mathrm{b}}$ & F$^{\mathrm{c}}$ & S$^{\mathrm{d}}$  & Counts$^{\mathrm{e}}$ & Counts  \\
ID                     &ID  &   &    &  & per sr$^{\mathrm{f}}$     \\
\hline
S2   & 13  & 220 & 1.20 & 27727 &  1.51 \\
S18  & 31  & 131 & 1.57 & 15584 &  2.00  \\
S6   & 36  & 108 & 1.84 & 26873 &  4.72  \\
S16  & 46f &  63 & 5.90 & 12398 & 10.42 \\   
--   & 46g &  39 & 3.47 &  8721 &  7.33  \\
--   & 59  &  45 & 0.66 &  7729 &  1.12 \\
--   & 67c & 105 & 5.06 & 15884 &  7.16 \\
S11  & 70  &  26 & 2.09 &  3489 &  2.93   \\
\hline
\end{tabular}
\begin{list}{}{}
\item[$^{\mathrm{a}}$] Sandage identification (Sandage \cite{sandage71}).
\item[$^{\mathrm{b}}$] HLG identification (Hodge et al. \cite{hodge90}).
\item[$^{\mathrm{c}}$] HLG published H$\alpha$ fluxes, in units of 
10$^{-15}$ \mbox{erg cm$^{-2}$ s$^{-1}$}.
\item[$^{\mathrm{d}}$] Surface brightness derived from HLG fluxes, in units 
of 10$^{-5}$ \mbox{erg cm$^{-2}$ s$^{-1}$ sr$^{-1}$}.
\item[$^{\mathrm{e}}$] Counts (sky subtracted) in our image integrated over
the areas defined by HLG.
\item[$^{\mathrm{f}}$] In units of 10$^{12}$ counts sr$^{-1}$.
\end{list}
\end{table*}

\begin{table*}
\caption[ ]{Measured parameters for \ion{H}{ii} regions in \object{IC 1613}}
\label{regiones}
\begin{tabular}{llllllllll}
\hline
Source$^{\mathrm{a}}$ &Sandage$^{\mathrm{b}}$ & HLG$^{\mathrm{c}}$ & D$^{\mathrm{d}}$ 
& $V_1$$^{\mathrm{e}}$ & $V_2$  &$\sigma_1$$^{\mathrm{f}}$ & $\sigma_2$  
& log F(H$\alpha$)$^{\mathrm{g}}$ & log L(H$\alpha$)$^{\mathrm{h}}$ \\
\hline
N1  & S2     &  13      & 127.6 & -248   &      & 20.0   &       & -12.41   &  37.34  \\
N2  &        & (25),30  & 53.3  & -242   &      & 18.8   &       & -13.50   &  36.25  \\ 
N3  &        & 26       & 78.1  & -254   &      & 31.4   &       & -13.28   &  36.47  \\
N4  &        & 27,(33)  & 47.2  & -242   &      & 26.5   &       & -13.72   &  36.03  \\
N5  &S12     & 29,(35)  & 80.5  & -247   &      & 21.2   &       & -12.83   &  36.91  \\
N6  &S18     & 31,32    & 99.0  & -247   &      & 21.2   &       & -12.89   &  36.86  \\
N7  &        & 34       & 34.2  & -244   &      & 21.2   &       & -13.53   &  36.22  \\
N8  &S6      & 36       & 57.5  & -240   &      & 12.0   &       & -12.89   &  36.86  \\
N9  &S3**    & 37       & 253.6 & -229   &      & 18.8   &       & -12.14   &  37.61  \\
N10 &        & 38       & 64.6  & -239   &      & 18.8   &       & -13.36   &  36.39  \\
N11 &S7      & 39,40    & 56.6  & -239   &      & 10.4   &       & -12.87   &  36.88  \\
N12 &        & 42,59    & 117.3 & -229   &      & 21.2   &       & -13.08   &  36.67  \\
N13 &        & (44)     & 29.3  & (-282) &      & (5.9)  &       & -14.74   &  35.01  \\
N14 &S17     & 45       & 90.8  & -236   &      & 18.8   &       & -12.79   &  36.96  \\
N15 & \bf{S15,S16}&\bf{46}& 138.3 & -244   &      & 22.2   &       & -12.14   &  37.61  \\
N16 &        & 47,57    & 34.7  & -236   &      & 17.6   &       & -13.27   &  36.48  \\
N17 &        & 48,58,61 & 60.1  & -235   &      & 16.3   &       & -13.09   &  36.66  \\
N18 & S8*    & 49       & 34.5  & -239   & -335 & 34.3   & 86.1  & -12.54   &  37.21  \\
N19 &        & 50,41    & 14.0  & -242   &      & 17.6   &       & -13.61   &  36.14  \\
N20 &S5      & 51,52,53 & 93.5  & -238   &      & 16.3   &       & -13.05   &  36.70  \\
N21 &\bf{S14}&\bf{55}   & 108.1 & -238   &      & 18.8   &       & -12.63   &  37.12  \\
N22 &        & 56       & 44.5  & -236   &      & 22.2   &       & -13.28   &  36.47 \\
N23 &        & (60)     & 29.1  & -242   &      & 22.2   &       & -14.51   &  35.24 \\
N24 &        & 63       & 26.9  & -240   &      & 24.4   &       & -13.94   &  35.81 \\
N25 &        & 64       & 76.3  & -237   &      & 24.4   &       & -12.87   &  36.88 \\
N26 &        & 66       & 184.5 & -238   &      & 21.2   &       & -12.46   &  37.29 \\
N27 &\bf{S10,S13}&\bf{67}& 198.1 & -248   & -185 & 22.2   & 18.9  & -11.98   &  37.77 \\
N28 &        &\bf{69}   & 42.2  & -228   &      & 29.5   &       & -13.63   &  36.12 \\
N29 & S11    & 70,73    & 42.7  & -235   &      & 22.2   &       & -13.53   &  36.22 \\
N30 &        &\bf{72}        & 99.0  & -228   &      & 32.4   &       & -13.06   &  36.69 \\
\hline
\end{tabular}
\begin{list}{}{}
\item[$^{\mathrm{a}}$] Identification used in this work.
\item[$^{\mathrm{b}}$] Sandage identification (Sandage \cite{sandage71}).
The regions marked in boldface are known superbubbles previously catalogued as 
\ion{H}{ii} regions. The asterisks indicate the regions hosting the supernova remnant 
(a single asterisk) and the Wolf-Rayet star (two asterisks).     
\item[$^{\mathrm{c}}$] HLG identification (Hodge et al. \cite{hodge90}). 
The regions marked in boldface are known superbubbles previously catalogued as 
\ion{H}{ii} regions. The regions quoted between parentheses are objects with uncertain
identification. 
\item[$^{\mathrm{d}}$] Equivalent diameter in parsecs (see text for definition).
\item[$^{\mathrm{e}}$] Peak velocity in units of km s$^{-1}$. Parentheses indicate
uncertain values. 
\item[$^{\mathrm{f}}$] Velocity dispersion in units of km s$^{-1}$. Parentheses 
indicate uncertain values. 
\item[$^{\mathrm{g}}$] Logarithm of the H$\alpha$ flux in units of erg cm$^{-2}$ s$^{-1}$.
\item[$^{\mathrm{h}}$] Logarithm of the H$\alpha$ luminosity in units of erg s$^{-1}$.
\end{list}
\end{table*}

\begin{table*}
\caption[ ]{Stellar content of the \ion{H}{ii} regions in \object{IC 1613}}
\label{stars}
\begin{tabular}{lllllll}
\hline
Source & Known & Related & Stellar & comments & Stellar & clusters
$^{\mathrm{e}}$ \\
      & stars$^{\mathrm{a}}$ & object $^{\mathrm{b}}$  & association
$^{\mathrm{c}}$ & & association $^{\mathrm{d}}$&  \\
\hline
N1   &      &     &          &          & H5  & c32, c41  \\
N2   &      &     &          &          & H10 &          \\
N3   &      &     & G10      &          & H11 &          \\
N4   &      &     & G11      &          &     &          \\
N5   & AM4  & O-B & G7       & embedded & H10 &          \\
N6   &      &     & G10      & center   & H11 &  c13     \\
N7   &      &     &          &          & H13 &          \\
N8   &      &     &          &          & H12 &          \\
N9   & AM6  & WO  &          &          &     &          \\
N10  &      &     & G11      & center   & H13 & c14      \\
N11  &      &     & G14      & embedded & H14 & c40,cl2  \\
N12  &      &     & G13      &          &     &          \\
N13  &      &     &          &          &     &          \\
N14  &      &     & G19      &          &     &          \\
N15  & AM5  & O-B & G16, G17 & embedded & H13 &cl1      \\
N16  &      &     & G14      & embedded & H14 &         \\
N17  &     &     & G18      & embedded & H14 &         \\
N18  & AM8  & SNR & G14      & embedded & H14 & c16     \\
N19  &      &     & G12,G14  & embedded & H14 &         \\
N20  &      &     & G15      & center   & H15 & c18     \\
N21  &      &     & G17, G21 & embedded & H13 & c15     \\
N22  &      &     &          &          &     &         \\
N23  &      &     &          &          &     &         \\
N24  &      &     & G24      & embedded & H17 &         \\
N25  &      &     & G23, G24 & embedded &     &         \\
N26  &      &     & G19      &          &     &         \\
N27  &      && G21, G25, G28 & embedded & H17 &         \\
N28  &      &     & G28      & embedded &     &         \\
N29  &      &     &          &          &     &         \\
N30  &      &     & G28      &          & H18 &         \\
\hline
\end{tabular}
\begin{list}{}{}
\item[$^{\mathrm{a}}$] Stars with known spectra: WR candidates from Armandroff \& Massey
 (\cite{armandroff85})
\item[$^{\mathrm{b}}$] Spectral types from  Azzopardi et al.
 (\cite{azzopardi88}).
\item[$^{\mathrm{c}}$] Associations determined  in Paper I.
\item[$^{\mathrm{d}}$] Associations determined by Hodge (1978).
\item[$^{\mathrm{e}}$] Stellar clusters determined in Hodge (1978) and in Paper I.
\end{list}
\end{table*}

\begin{table*}
\caption[ ]{Catalogue of Superbubbles in the galaxy IC 1613}
\label{cat_shells}
\begin{tabular}{llllllll}
\hline
Source& ID$_{1}$$^{\mathrm{a}}$ & axis$^{\mathrm{c}}$ & $V_1$ & $V_2$ & log
F(H$\alpha$)$^{\mathrm{d}}$ & log L(H$\alpha$)$^{\mathrm{e}}$ & Comments  \\
      & (ID$_{2})$$^{\mathrm{b}}$ & $\arcsec \times \arcsec$ & km s$^{-1}$ 
& km s$^{-1}$ & & &\\ 
\hline
R1 & GS1a & 39 $\times$ 39 & -216 & -274   & -12.23   & 37.51 & High [SII] emission \\
   & (S13)              & & & & & & \\

R2 & GS1b & 49 $\times$ 33 & -244 & -147   & -11.98   & 37.77 & High [SII] emission \\
   & (S10)               & & & & & & \\

R3 &      & 73 $\times$ 59 & -238 & -179   & -12.23   & 37.52 & Faint [SII] emission \\
   & (HIV) (HHG 2)      & & & & & &	\\
R4 & GS2  & 66 $\times$ 50 & -277 & -213   & -12.18   & 37.57 & Ring of HII regions \\
   & (HHG 1) &            &      &       &	               & \\

R5 & GS4  & 33 $\times$ 30 & -228 & -153   & -12.10   & 37.65 & High [SII] emission\\
   & (S15)                 & & & & & & \\

R6 & GS5  & 64 $\times$ 52 & -240 & -190   & -12.64   & 37.11 & Faint [SII] emission \\
   & (S16+HIII) (HHG 4) & & & & & & \\

R7 & GS6  & 47 $\times$ 38 & -234 & (-174) & -12.72   & 37.03 &  \\
   & (HIV) (HHG 5)      & & & & & & \\

R8 & --   & 63 $\times$ 49 & -259 & -208   & -13.08   & 36.67 & \\
   & (HVII) (HHG 3)     & & & & & & \\

R9 & --   & 90 $\times$ 90 & -237 &  -317  & (-12.31) & 37.44 & Ring of HII regions\\
   & (H5+S5+S7+S8)         & & & & & &  \\
R10 & --  & 37 $\times$ 37 & -229 & (-182) & -12.71   & 37.04 & [SII] detected \\
    & (S6(contained))      & & & & &              & S8 SNR$^{\mathrm{f}}$  \\
R11 & --  & 77 $\times$ 77 & -205 &   --    & -13.19   & 36.56 & Very faint \\

R12 & --               & 108 $\times$ 61 & -210  & -138 & -12.69 & 37.06 &  \\
    & (HII(contained)) & & & & & & \\
R13 & GS3 & 37 $\times$ 24 & -217 & -296 & -12.62 & 37.13 & Expansion better \\
    & (S12+S14)        & & & &                        & & seen in [SII] \\
 
R14 & --      & 85 $\times$ 77  & -234  & --    & -13.39  & 36.36 & Very faint	\\
    & (HXII)  & & & &	               & & \\

R15 & --     & 45 $\times$ 35  & -254  & -199 & - 13.20 & 36.55 & Very faint	\\
    & (HIX)  & & & &	               & & \\

R16 & --   & 73 $\times$ 64  & -210  &   --  & -12.89  & 36.85 & Very faint	\\
    & (D1) & & & &	               & & \\

R17 & --             & 108 $\times$ 59 & -215  & --    & -13.43  & 36.32 &	\\
    & (HI contained) & & & &	               & & \\
\hline
\end{tabular}
\begin{list}{}{}
\item[$^{\mathrm{a}}$] Meaburn et al. (\cite{meaburn88}) identification.
\item[$^{\mathrm{b}}$] Price et al. (\cite{price90}) or HHG for Hunter et al.
(\cite{hunter93}) identifications.
\item[$^{\mathrm{c}}$] Dimensions of the superbubble semiaxis in arcsecond 
units (see details in section 4.2).
\item[$^{\mathrm{d}}$] Flux logarithm in units of erg cm$^{-2}$ s$^{-1}$.
\item[$^{\mathrm{e}}$] Luminosity logarithm in units of erg s$^{-1}$.
\item[$^{\mathrm{f}}$] S8 SNR projected at the boundary of this superbubble.
 \end{list}
\end{table*}

\begin{table*}
\caption[ ]{Stellar Content of the Superbubbles in the galaxy \object{IC 1613}}
\label{stars_shell}
\begin{tabular}{llllllc}
\hline
Source & Paper I      &  N$_{B}$ & Age   & Known &  Blue & $\Gamma$  \\
       & Hodge (1978) &          & (Myr) & stars &  SG   &           \\
\hline
R1 & part of G28 &  3  & 20 & no  & no & -2.4 +/- 0.3 \\
   & H18         &     &    &     &    &              \\

R2 & G25, part of G28 & 17 & 8 & no  & one    & -1.6 +/- 0.2   \\
   & H17              &    &   &     &  (B0)  &                \\

R3 & G29 & 8 & 20 & no & no & -2.1 +/- 0.1 \\
   & H19 &   &    &    &    &              \\

R4  & parts of G16, G17, G21, G24, G25 & 37 & 8--10 & no & no & -1.6 +/- 0.2 \\
    &  H13, H17 (parts of)             &    &      &    &    &              \\

R5  & part of G16    & 7  & 10   & one       & one   & -1.7 +/- 0.2 \\
    &  H13 (part of) &    &      & (OBem(?)) & (S13) &              \\

R6 & G5, G7, G11         &  36 & 5 & one  & two            & -1.3 +/- 0.3 \\
   &  H10, H13 (part of) &     &   & (OB) &  (cB1.5, A0Ia) &              \\

R7 & G9, part of G11 & 16 & 10 & no  & no  & -1.7 +/- 0.2 \\
   &  H13 (part of)  &    &    &     &     &              \\

R8  & parts of , G9, G10  & 5 & -- & no &  one   & --  \\
    &                     &   &    &    & (B2I)  &     \\

R9 & G12, G14, G15, G18, part of G6, G13 & 106 & 5--10 & no & no & -2.0 +/- 0.2 \\
   &  H14, H15                           &     &      &    &    &              \\

R10  & G8, part of G12 & 7 & 5 & no  & one & -1.1 +/- 0.2  \\
     &part of  H14     &   &   &     &     &               \\

R11 &-- & -- & -- &-- & -- & --  \\
    &   &    &    &   &    &    \\

R12 & -- & -- & -- & -- & --  &  \\
    &    &    &    &    &     &  \\

R13  & part of G21 & 9 & 8 &  no  & no  & -1.6 +/- 0.2  \\
     & part of H17 &   &   &      &     &               \\

R14 & --      & 21  & 16 & no & no & -- \\
    &  H6, H5 &     &    &    &    &    \\

R15 & --   & 36 & 31 & no & no & -- \\
    & H4   &    &    &    &    &    \\

R16 & -- & 8 & 24  & no & no & -- \\
    & H8 &   &     &    &    &    \\

R17 & --  &  25 & 11 & no & no & -- \\
    &  H9 &     &    &    &    &    \\
\hline
\end{tabular}
\end{table*}

\begin{table*}
\caption[ ]{Physical Parameters of the Superbubbles in the galaxy 
\object{IC 1613}}
\label{parameters_shells}
\begin{tabular}{lllccl}
\hline
Source& R & n$_{e}$(rms) & $V_{EXP}$ & K.E$^{\mathrm{a}}$ & t$^{\mathrm{b}}$ \\
      & pc & cm$^{-3}$ & km s$^{-1}$ &  ergs              & Myr \\
\hline
R1  & 70  & 3.8   &  29  & 1.88   & 1.2   \\
R2  & 74  & 4.9   &  49  & 8.07   & 0.7   \\
R3  & 119 & 1.8   &  30  & 4.64   & 1.9   \\
R4  & 105 & 2.2   &  32  & 4.49   & 1.6   \\
R5  & 55  & 6.3   & 38   & 2.56   & 0.7   \\
R6  & 105 & 1.4   & 25   & 1.70   & 2.0   \\
R7  & 77  & 1.9   & (30) & (1.34) & (1.2) \\
R8  & 101 & 0.8   & 26   & 1.01   & 1.9   \\
R9  & 162 & (1.1) & --   & --     & --    \\
R10 & 66  & 2.4   & (24) & (0.69) & (1.3) \\
R11 & 138 & 0.4   &  --  & --     & --    \\
R12 & 152 & 0.7   & 36   & 5.75   & 2.0   \\
R13 & 55  & 3.6   &  32  & 1.04   & 0.8   \\
R14 & 146 & 0.3   & --   & --     & --    \\
R15 & 72  & 1.3   & 28   & 0.64   & 1.2   \\
R16 & 123 & 0.7   & --   & --     & --    \\
R17 & 150 & 0.3   & --   & --     & --    \\
\hline
\end{tabular}
\begin{list}{}{}
\item[$^{\mathrm{a}}$] Mechanical energy in units of 10$^{50}$ ergs. See section 5.2
for details.
\item[$^{\mathrm{b}}$] Dynamical age in units of Myrs. See section 5.2 for details.
 \end{list}

\end{table*}

\begin{table*}
\caption[ ]{Comparison between the predicted wind power and the detected
blue stars for the superbubbles of \object{IC 1613}}
\label{wind_power}
\begin{tabular}{lll}
\hline
Source& L$_{W}$$^{\mathrm{a}}$ & N$_{B}$$^{\mathrm{b}}$ \\
\hline
R1  & 36  & 3   \\
R2  & 253 & 17  \\
R3  & 55  & 8   \\
R4  & 64  & 37  \\
R5  & 84  & 7   \\
R6  & 19  & 36  \\
R7  & 24  & 16  \\
R8  & 11  & 5   \\
R9  & --  & 106 \\
R10 & 12  & 7   \\
R11 & --  & --  \\
R12 & 60  & --  \\
R13 & 29  & 9   \\
R14 & --  & 21  \\
R15 & 12  & 36  \\
R16 & --  & 8   \\
R17 & --  & 25  \\

\hline
\end{tabular}
\begin{list}{}{}
\item[$^{\mathrm{a}}$] Predicted wind power in units of 10$^{36}$ ergs 
s$^{-1}$. See section 5.4 for details.
\item[$^{\mathrm{b}}$] Observed blue stars. See section 5.4 for details.
\end{list}

\end{table*}

\end{document}